\documentclass[useAMS,usenatbib]{mn2e}
\usepackage{amsmath}
\usepackage{times}
\usepackage{ctable,longtable}
\usepackage{graphicx}
\renewcommand{\thefootnote}{\alph{footnote}}
\usepackage{footnote}
\makesavenoteenv{tabular}
\long\def\symbolfootnote[#1]#2{\begingroup%
\def\thefootnote{\fnsymbol{footnote}}\footnote[#1]{#2}\endgroup}

\title[\textit{Herschel} and \textit{JCMT} observations of the early-type dwarf galaxy NGC\,205]{\textit{Herschel}\symbolfootnote[1]{} and JCMT observations of the early-type dwarf galaxy NGC\,205}

\author[I. De Looze et al.]{
I. De Looze,$^1$ M. Baes,$^1$ T.~J. Parkin,$^2$ C.~D. Wilson,$^2$ G.~J. Bendo,$^3$ M. Boquien,$^4$
\newauthor 
 A. Boselli,$^4$ A. Cooray,$^5$  D. Cormier,$^6$ J. Fritz,$^1$ F. Galliano,$^6$ W. Gear,$^7$  G. Gentile,$^1$
\newauthor
 V. Lebouteiller,$^6$ S.~C. Madden,$^6$  H. Roussel,$^8$  M. Sauvage,$^6$ M.~W.~L. Smith,$^7$ 
\newauthor 
 L. Spinoglio,$^9$ J. Verstappen,$^1$ L. Young$^{10}$\\
$^1$Sterrenkundig Observatorium, Universiteit Gent, Krijgslaan 281 S9, B-9000 Gent, Belgium \\
$^2$ Dept. of Physics \& Astronomy, McMaster University, Hamilton, Ontario, L8S 4M1, Canada \\
$^3$UK ALMA Regional Centre Node, Jodrell Bank Centre for Astrophysics, School of Physics and Astronomy, \\
University of Manchester, Oxford Road, Manchester M13 9PL, United Kingdom \\
$^4$ Laboratoire d'Astrophysique de Marseille - LAM, Universit{\'e} d'Aix-Marseille \& CNRS, UMR7326, 38 rue F. Joliot-Curie, 13388 Marseille Cedex 13, France \\
$^5$ Department of Physics \& Astronomy, University of California, Irvine, CA 92697, USA \\
$^6$ Laboratoire AIM, CEA/DSM- CNRS - Universit\'e Paris Diderot, Irfu/Service d'Astrophysique, 91191 Gif sur Yvette, France \\
$^7$ School of Physics and Astronomy, Cardiff University, Queens Buildings The Parade, Cardiff CF24 3A A, UK \\ 
$^8$ Institut dÕAstrophysique de Paris, and Universit{'e} Pierre et Marie Curie (UPMC), UMR 7095 CNRS, 98 bis boulevard Arago, 75014 Paris \\
$^9$ Istituto di Fisica dello Spazio Interplanetario, INAF, Via del Fosso del Cavaliere 100, I-00133 Roma, Italy \\
$^{10}$ Physics Department, New Mexico Institute of Mining and Technology, Socorro, NM 87801, USA 
}

\begin{document}

\pagerange{\pageref{firstpage}--\pageref{lastpage}} \pubyear{2012}
\maketitle

\label{firstpage}

\begin{abstract}
We present \textit{Herschel} dust continuum, James Clerk Maxwell Telescope CO(3-2) observations and a search for [C{\sc{ii}}] 158 $\mu$m and [O{\sc{i}}] 63 $\mu$m spectral line emission for the brightest early-type dwarf satellite of Andromeda, NGC\,205.
While direct gas measurements ($M_{g}$ $\sim$ 1.5 $\times$ 10$^6$ M$_{\odot}$, H{\sc{i}} + CO(1-0)) have proven to be inconsistent with theoretical predictions of the current gas reservoir in NGC\,205 ($>$ 10$^7$ M$_{\odot}$), we revise the missing interstellar medium mass problem based on new gas mass estimates (CO(3-2), [C{\sc{ii}}], [O{\sc{i}}]) and indirect measurements of the interstellar medium content through dust continuum emission.  

Based on \textit{Herschel} observations, covering a wide wavelength range from 70 to 500 $\mu$m, we are able to probe the entire dust content in NGC\,205 ($M_{\text{d}}$ $\sim$ 1.1-1.8 $\times$ 10$^4$ M$_{\odot}$ at $T_{d}$ $\sim$ 18-22 K) and rule out the presence of a massive cold  dust component ($M_{\text{d}}$ $\sim$ 5 $\times$ 10$^5$ M$_{\odot}$, $T_{d}$ $\sim$ 12 K), which was suggested based on millimeter observations from the inner 18.4$\arcsec$. 
Assuming a reasonable gas-to-dust ratio of $\sim$ 400, the dust mass in NGC\,205 translates into a gas mass $M_{g}$ $\sim$ 4-7 $\times$ 10$^{6}$ M$_{\odot}$.
The non-detection of [O{\sc{i}}] and the low $L_{\text{[CII]}}$-to-$L_{\text{CO(1-0)}}$ line intensity ratio ($\sim$ 1850) imply that the molecular gas phase is well traced by CO molecules in NGC\,205.
We estimate an atomic gas mass of 1.5 $\times$ 10$^{4}$ M$_{\odot}$ associated with the [C{\sc{ii}}] emitting PDR regions in NGC\,205.
From the partial CO(3-2) map of the northern region in NGC\,205, we derive a molecular gas mass of $M_{\text{H}_{2}}$ $\sim$ 1.3 $\times$ 10$^5$ M$_{\odot}$.
Upon comparison with the molecular gas mass estimated from CO(1-0) observations ($M_{\text{H}_{2}}$ $\sim$ 6.9 $\times$ 10$^5$ M$_{\odot}$), we find most of the H$_{2}$ gas in NGC\,205 to be locked in diffuse regions of low density and/or temperature, characteristic for an interstellar medium with little star formation activity. 

New total gas mass estimates from \textit{Herschel} dust continuum (4-7 $\times$ 10$^{6}$ M$_{\odot}$), \textit{Herschel} [C{\sc{ii}}] line spectroscopic mapping (1.5 $\times$ 10$^{4}$ M$_{\odot}$) and James Clerk Maxwell Telescope CO(3-2) observations (7 $\times$ 10$^{5}$ M$_{\odot}$), including the H{\sc{i}} mass ($M_{\text{HI}}$ $\sim$ 4.0 $\times$ 10$^{5}$ M$_{\odot}$) and a correction for heavier elements, confirm the deficiency of the interstellar medium (gas+dust) in the inner regions of NGC\,205, which is predicted to contain at least $>$ 10$^{7}$ M$_{\odot}$ of gas if we assume a reasonable star formation efficiency of 10$\%$ and account for the mass return from planetary nebulae.
In an attempt to explain the missing interstellar medium mass problem, we claim that efficient supernova feedback capable of expelling gas from the inner, star-forming regions to the outer regions and/or tidal interactions with M31 stripping the gas component from the galaxy provide the best explanation for the removal of a significant amount of gas and dust from NGC\,205. 
\end{abstract}


\section{Introduction}

\footnotetext[1]{\textit{Herschel} is an ESA space observatory with science instruments provided by European-led Principal Investigator consortia and with important participation from NASA.}

Eighty-five percent of all galaxies are located outside galaxy clusters, among which half reside in groups \citep{2005AJ....129..178K}. Therefore, studying group environments is of great importance to learn more about the habitats for an important fraction of galaxies. The Local Group is of particular interest for studies of low surface brightness galaxies, since galaxies at the low luminosity end often remain undetected in more distant group structures. Moreover, the Local Group allows probing the interstellar medium (ISM) of its residents at high resolution and, at the same time, offers a wealth of ancillary data being one of the best studied areas on the sky. Studying the properties of the ISM and metal-enrichment in metal-poor dwarf galaxies offers a promising way to learn more about the conditions in the early Universe and the evolution of dwarf galaxies throughout the history of the universe.

Among the low surface brightness galaxies in the Local Group, the dwarf satellite NGC\,205 ($\sim$ 824 kpc, \citealt{2005MNRAS.356..979M}) of the Andromeda galaxy is of particular importance due to its relatively low metal abundance (Z $\sim$ 0.13 Z$_{\odot}$, \citealt{2008ApJ...684.1190R}), interesting star formation history and indications of a tidal encounter with its massive companion M31. Although photometrically classified as a dwarf elliptical galaxy \citep{1991rc3..book.....D}, the formation processes for NGC\,205 seem more closely related to the dwarf spheroidal galaxy population (transformed from late-type galaxies through internal and environmental processes) rather than to merger remnants, thought to be the main driver for the formation of genuine ellipticals \citep{2009ApJS..182..216K}.
The star formation history in NGC\,205 has been studied extensively \citep{1951POMic..10....7B, 1973ApJ...182..671H, 1995ApJ...438..680B, 2003ApJ...597..289D}.
An old stellar population (10 Gyr, \citealt{1990A&A...228...23B}) dominates the overall stellar content of the dwarf galaxy and a plume of bright blue star clusters in the central region of NGC\,205 was already identified $\sim$ 60 years ago \citep{1951POMic..10....7B, 1973ApJ...182..671H}. 

\subsection{Theoretical gas mass predictions}
\label{Theory.sec}
Combining observations of this young stellar population with adequate model assumptions has provided several independent theoretical predictions of the current gas content in NGC\,205.
Such predictions need to account for both the left-over gas reservoir after an epoch of star formation activity and the build-up of gas returned to the ISM by the evolved stellar population since the last starburst episode.

The left-over gas reservoir is probed through observations of the young stellar population providing an estimate of the total gas mass consumed during the last epoch of star formation. Usually, it is assumed that a star formation efficiency of 10$\%$ is reasonable (e.g. \citealt{2011arXiv1111.1925M}, \citealt{Boylan}).
From observations of both the nucleus and a region about 1$\arcmin$ north of the nucleus with the International Ultraviolet Explorer (IUE) within an aperture of 10$\arcsec$ $\times$ 20$\arcsec$, 
 \citet{1990ApJ...364...87W} report a mass of young ($\sim$ 10 Myr old) stars $M_{\star}$ $\sim$ 7 $\times$ 10$^5$ M$_{\odot}$ in NGC\,205, which increases to $M_{\star}$ $\sim$1.4 $\times$ 10$^6$ M$_{\odot}$ when extrapolating to the whole galaxy based on the stellar light contribution from OB stars \citep{1990ApJ...364...87W} and also taking stellar masses $<$ 1 M$_{\odot}$ into account \citep{2006ApJ...646..929M}. 
According to \citet{1995ApJ...438..680B} NGC\,205 was the host of a starburst starting $\sim$ 500 Myr ago involving a total stellar burst mass of $M_{\star}$ $\sim$ 5.3 $\times$ 10$^7$ M$_{\odot}$, which is considerably higher than the estimate in \citet{1990ApJ...364...87W}.
More recently, \citet{2009A&A...502L...9M} estimate $M_{\star}$ $\sim$ 1.9 $\times$ 10$^5$ M$_{\odot}$ of stars to be produced between $\sim$ 62 Myr and $\sim$ 335 Myr ago from observations of the nuclear 29$\arcsec$ $\times$ 26$\arcsec$ region of NGC\,205 with the Advanced Camera for Survey (ACS) on board the Hubble Space Telescope (HST), when assuming a $\Lambda$CDM cosmological model.
Although the estimated burst mass in \citet{2009A&A...502L...9M} also includes lower mass stars ($M_{\star}$  $<$ 1 M$_{\odot}$), their value should be considered a lower limit of the total burst mass, since only a limited period (from 62 to 335 Myr ago) in the star formation history of NGC\,205 is analyzed and the observed area only corresponds to part of the region where the most recent star formation epoch took place.  

The total amount of gas returned to the ISM by planetary nebulae is predicted to be $\sim$ 1.8 $\times$ 10$^6$ M$_{\odot}$ \citep{1998ApJ...499..209W}, following the prescriptions in \citet{1976ApJ...204..365F} and assuming a time lapse of $\sim$ 500 Myr since the trigger of the last star formation activity. With the lower limit for the stellar burst estimates relying on 10 Myr old stars \citep{1990ApJ...364...87W}, the corresponding mass returned to the ISM since the formation of those young stellar objects can be predicted in a similar way ($\sim$ 3.6 $\times$ 10$^4$ M$_{\odot}$).
We only account for the mass loss from planetary nebulae since the mass lost from more massive stars is considered negligible due to their lower mass loss rate and shorter lifetime.
Considering that the estimated burst mass critically depends on the model assumptions and is often biased by the sensitivity and coverage of the observations, we calculate a total burst mass during the last episode of star formation in the range 1.4 $\times$ 10$^6$ M$_{\odot}$ $\leq$ $M_{\star}$ $\leq$ 5.3 $\times$ 10$^7$ M$_{\odot}$, where the lower and upper limits correspond to stellar burst mass predictions from \citet{1990ApJ...364...87W} and \citet{1995ApJ...438..680B}, respectively. A burst mass of 1.4 $\times$ 10$^6$ M$_{\odot}$ would predict that the initial gas reservoir before the star formation (SF) episode was $M_{g}$ $\sim$ 1.4 $\times$ 10$^7$ M$_{\odot}$ for a star formation efficiency (SFE) of $\sim$ 10$\%$. Subtracting the 10$\%$ of gas consumed into stars results in the left-over gas reservoir $M_{\text{g}}$ $\sim$ 1.3 $\times$ 10$^7$ M$_{\odot}$ after the star formation epoch. When combining the left-over gas reservoir (i.e. the majority of the ISM mass) and mass loss by planetary nebulae (depending on the assumed time lapse), we estimate a current gas content for NGC\,205 ranging between 1.3 $\times$ 10$^7$ M$_{\odot}$ $\leq$ $M_{\text{g}}$ $\leq$ 4.8 $\times$ 10$^8$ M$_{\odot}$.

\subsection{Observations of the ISM content}
\label{gasobs.sec}
Up to now, the total gas mass in NGC\,205 was estimated from H{\sc{i}}, CO(1-0) and dust continuum observations.
A total H{\sc{i}} mass of 4.0 $\times$ 10$^5$ M$_{\odot}$, scaled to a distance $D$ = 824 kpc for NGC\,205, is reported in \citet{1997ApJ...476..127Y} based on VLA observations covering the whole galaxy. 
\citet{1998ApJ...499..209W} detect CO(1-0) and CO(2-1) emission above the 3$\sigma$ level from 3 and 4 positions, respectively, across the plane of NGC\,205. 
Although a partial beam overlap occurs for the CO(1-0) observations (see red circles in Figure \ref{hicontourmap}), the covered area in the CO(1-0) observations is 7 times larger than for the second CO(2-1) transition and, thus, provides a better estimate of the molecular gas content in NGC\,205. 
While \citet{1998ApJ...499..209W} assumed a close to solar metallicity (implying a CO-to-$H_{\text{2}}$ conversion factor $X_{\text{CO}}$ $=$ 2.3 $\times$ 10$^{20}$ cm$^{-2}$ (K km s$^{-1}$)$^{-1}$, \citealt{1988A&A...207....1S}), we apply a conversion factor of 6.6 $\times$ 10$^{20}$ cm$^{-2}$ (K km s$^{-1}$)$^{-1}$, determined from the expression reported in \citet{2002A&A...384...33B} relating the $X_{\text{CO}}$ factor to the oxygen abundance in a galaxy. This value for the $X_{\text{CO}}$ factor is based on a metal abundance of $Z$ $\sim$ 0.3 Z$_{\odot}$ in the inner regions of NGC\,205.
Whereas the earlier reported metallicity value (Z $\sim$ 0.13 Z$_{\odot}$) was obtained from averaging the oxygen abundances for 13 planetary nebulae in NGC\,205 \citep{2008ApJ...684.1190R} and thus refers to the whole galaxy, \citet{2006MNRAS.372.1259S}  determined a mean metallicity [Z/H] $\sim$ - 0.5 $\pm$ 0.2 for the central regions in NGC\,205 from Lick indices.  Considering that the last episode of star formation mainly occurred in the inner most regions, this gradient in metallicity and/or age is not surprising.  
A similar central increase in colour and metallicity has been noticed in several early-type dwarf galaxies in the Fornax cluster \citep{2009MNRAS.396.2133K} and a population of dEs with central blue cores has been observed in the Virgo cluster \citep{Lisker2006_2}.

Additional CO(1-0) line emission was detected from a small area (beamsize $\sim$ 21$\arcsec$) observed in the south of NGC\,205 \citep{1996ApJ...464L..59Y} (see green circle in Figure \ref{hicontourmap}). 
Combining those CO(1-0) observations, we obtain an estimate of $M_{\text{H}_{2}}$ $=$ 6.9 $\times$ 10$^5$ M$_{\odot}$. This value for the molecular gas mass is however only a lower limit of the total H$_{2}$ mass, since the southern part of the galaxy is poorly covered by current CO(1-0) observations.
Combining both H{\sc{i}} and CO(1-0) observations, which were scaled by a factor of 1.4 to include helium, we derive a total gas mass $M_{g}$ $\sim$ 1.5 $\times$ 10$^{6}$ M$_{\odot}$ for NGC\,205, which is about one order of magnitude lower than the more modest theoretical predictions for the gas content. This deficiency in the gas content of NGC\,205 is often referred to as the problem of the ``missing ISM mass" \citep{1998ApJ...499..209W}.

Although gas observations directly probe the ISM component of interest, the dependence of the $X_{\text{CO}}$ conversion factor on the metallicity \citep{1995ApJ...448L..97W,2008ApJ...686..948B} and density of the gas \citep{2011MNRAS.412.1686S}, and the optical thickness of the CO(1-0) line introduce an uncertainty on the estimate of the total gas mass. In particular for metal-poor galaxies the molecular gas phase could be poorly traced by CO \citep{1995ApJ...448L..97W, 1997ApJ...483..200M, 2005ApJ...625..763L, 2007ApJ...658.1027L}.  
An alternative and promising method to measure the ISM mass in galaxies is to use observations of the continuum emission from dust (\citealt{1983QJRAS..24..267H, 1993A&A...279L..37G, 1995A&A...298L..29G, 2002A&A...384...33B, 2002MNRAS.335..753J,2010A&A...518L..62E}, Eales et al.\,in prep.).

Dust emission from NGC\,205 was first detected with IRAS \citep{1988ApJS...68...91R}. NGC\,205 was also the first early-type dwarf galaxy detected at millimeter wavelengths \citep{1991ApJ...374L..17F}.  Based on these \textit{IRAS} data and 1.1 mm observations for the central 18$\arcsec$ (21 $\pm$ 5 mJy) in NGC\,205, \citet{1991ApJ...374L..17F} estimated a dust mass of $M_{\text{d}}$ $\sim$ 3 $\times$ 10$^{3}$ M$_{\odot}$ at a temperature of $\sim$ 19K. Using ISO observations, \citet{1998A&A...337L...1H} obtained 
a total dust mass estimate of 4.9 $\times$ 10$^3$ M$_{\odot}$ at a temperature of $\sim$ 20 K. 
Recently, \citet{2006ApJ...646..929M} probed the dust emission from NGC\,205 with \textit{Spitzer} and found a dust mass in the range $M_{\text{d}}$ =  3.2-6.1 $\times$ 10$^{4}$ M$_{\odot}$ at a temperature of $\sim$ 18 K. Taking the 1.1 mm observation of the core region into account, \citet{2006ApJ...646..929M} found a dust component in the central regions at a temperature of $T_{d}$ $\sim$ 11.6, sixteen times more massive, suggesting that a substantial amount of cold dust might be overlooked if one only takes IRAS, ISO and Spitzer observations into consideration. 
Unfortunately only the central region was observed at 1.1 mm, and it could not be investigated whether such a putative cold dust component is present over the entire galaxy.
Under the assumption that the colder dust is not only distributed in the central region, but is abundantly present in the entire galaxy, they estimated a total gas mass of 5 $\times$ 10$^{7}$ M$_{\odot}$ for a gas-to-dust ratio of 100. 
To probe this cold dust component, we need observations at wavelengths longwards of 160 $\mu$m (e.g. \citealt{2010A&A...518L..89G}, \citealt{2011A&A...532A..56G}). 
Longer wavelength data also allow us to constrain the Rayleigh-Jeans side of the dust SED, from which a more robust temperature estimate for our SED model can be obtained. 
Probing this cold dust component is now possible with the \textit{Herschel} Space Telescope \citep{2010A&A...518L...1P}, covering a wavelength range from 70 up to 500 $\mu$m. 
Recent \textit{Herschel} observations of nearby galaxies have demonstrated the presence of significant amounts of cold dust (e.g. \citealt{2010A&A...518L..65B,2010A&A...518L..51S,2011arXiv1109.0237B,2011AJ....142..111B}; Fritz et al., subm.).

In this paper, we present \textit{Herschel} observations for NGC\,205 taken in the frame of the VNGS and HELGA projects, with the aim of making an inventory of all the dust in NGC\,205. Furthermore, we report new gas mass measurements from James Clerk Maxwell Telescope (JCMT) CO(3-2) observations and Herschel [C{\sc{ii}}] 158 $\mu$m and [O{\sc{i}}] 63 $\mu$m line spectroscopic mapping. From those new dust and gas mass estimates, we are able to revise the ``missing ISM" problem in NGC\,205.
In Section~{\ref{Observations.sec}}, the data and observing strategy from \textit{Herschel} and JCMT observations are discussed. The corresponding data reduction procedures are outlined and a brief overview of the ancillary dataset is given. 
Section \ref{Results.sec} discusses the spatial distribution of gas and dust in NGC\,205 ($\S$ \ref{Distribution.sec}), the global flux measurements ($\S$ \ref{Global.sec}) and the basic principles and results of the SED fitting procedure ($\S$ \ref{Dust.sec}).
The JCMT CO(3-2) and PACS spectroscopy data are analyzed in Section \ref{Gas.sec} and new estimates for the gas mass are determined. 
Section~{\ref{Discussion.sec}} reanalyses the missing ISM mass problem in NGC\,205 ($\S$ \ref{Problem.sec}), discusses the implications of our results for the SF conditions and properties of the ISM in the galaxy ($\S$ \ref{Explanation.sec}) and makes a comparison with the other early-type dwarf companions of Andromeda  ($\S$ \ref{Compare.sec}). Finally, Section \ref{Conclusions.sec} summarizes our conclusions.

\section{Observations and data reduction} 
\label{Observations.sec}   
\subsection{PACS Photometry}
\label{PACS.sec}
We use data for NGC\,205 taken as part of two \textit{Herschel} Guaranteed Time Projects: the Very Nearby Galaxy Survey (VNGS, PI: C. Wilson) and the \textit{Herschel} Exploitation of Local Galaxy Andromeda (HELGA, PI: J. Fritz).

From the VNGS, we obtained PACS photometry \citep{2010A&A...518L...2P} at 70 and 160 $\mu$m (ObsID 1342188692, 1342188693 ) and SPIRE photometry \citep{2010A&A...518L...3G} at 250, 350 and 500 $\mu$m.  
PACS data were observed on the 29th of December 2009 and cover an area of 1$^{\circ}$ by 1$^{\circ}$ centered on NGC\,205. This area was observed in nominal and orthogonal scan direction with four repetitions at a medium scan speed (20$\arcsec$/s).

The main scientific objective of HELGA (Fritz et al.\,subm.) focuses on dust in the extreme outskirts of the Andromeda Galaxy. Thanks to its large survey area, NGC\,205 was covered in the field of two overlapping scan observations. HELGA observations were performed in the parallel fast scan mapping mode (60$\arcsec$/s), obtaining PACS 100, 160 $\mu$m and SPIRE 250, 350 and 500 $\mu$m photometry (ObsID 1342211294, 1342211309, 1342211319, 1342213207).
 
To reduce the PACS data, we used version 13 of the \textit{Scanamorphos} (Roussel et al.\,in prep., \footnote{\it http://www2.iap.fr/users/roussel/herschel}) map making technique. Before applying \textit{Scanamorphos} to the level 1 data, the raw data were pre-processed in \textit{Herschel} Interactive Processing Environment (HIPE, \citealt{2010ASPC..434..139O}) version HIPE 6.0.1196. 
Due to the different observing set-ups, the depth of the PACS observations is inhomogeneous among the different wavebands.
For PACS observations at 70 and 100 $\mu$m, we only have data available from one survey (either VNGS or HELGA, respectively).
This dataset was finally reduced to obtain maps with a pixel size of 2$\arcsec$. The FWHM of the PACS beams are $\sim$ 6$\arcsec$ and $\sim$ 7$\arcsec$$\times$13$\arcsec$ at 70 and 100 $\mu$m, respectively. Due to the lower level of redundancy and the fast scan speed, the PACS 100 $\mu$m waveband was observed in the least favorable conditions, resulting in the largest uncertainty values and interference patterns affecting the observations (see Figure \ref{vngsmaps}, second panel on the top row).
In the red filter (PACS 160 $\mu$m), NGC\,205 was covered by both VNGS and HELGA in medium and fast scan speed, resulting in a FWHM for the PACS beam of  $\sim$ 12$\arcsec$ and $\sim$ 12$\arcsec$$\times$16$\arcsec$, respectively.
Our final photometry map combines data from both \textit{Herschel} projects at this overlapping wavelength with the aim of increasing the signal to noise ratio. 
The unmatched scan speeds prevent reducing both observations simultaneously, since the drift correction is calculated over a certain stability length which depends on the scan speed of the observation.  
Therefore, both datasets were reduced individually in \textit{Scanamorphos}, using the same astrometry for the final maps. Before combining both maps we convolved them to the same resolution of the PSF, to avoid issues with the different beam sizes in both observations.
Finally, the separately reduced and convolved VNGS and HELGA maps at 160 $\mu$m were combined into one single map in IRAF, with the \texttt{imcombine} task.
The images were produced with the latest version of the PACS calibration files (version 26) and divided by the appropriate colour correction factors (1.016, 1.034 and 1.075 at 70, 100 and 160 $\mu$m, see \citealt{Muller2011} for a power-law spectrum with $\beta$ $=$ 2).
The background was also subtracted from the final PACS images. An estimate for the background was obtained from averaging the background flux within 100 random apertures (diameter $=$ 4$\times$FWHM). 
Random apertures were selected within an annulus centered on NGC\,205 with inner radius of 5$\arcmin$ and outer radius of 20$\arcmin$, avoiding regions with bright emission from background sources and M31.
Once the random background apertures were selected, those same positions for aperture photometry were applied at all wavelengths.

\subsection{SPIRE Photometry}
\label{SPIRE.sec}
SPIRE data were observed on the 27th of December 2009 (ObsID 1342188661), obtaining two repetitions of nominal and orthogonal scans at medium scan speed (30$\arcsec$/s).
For all SPIRE bands, datasets from both VNGS and HELGA projects were available and combined into one frame.
In the same way as for \textit{Scanamorphos}, the continuous temperature variations are different for VNGS and HELGA observations. Therefore, the corresponding data are reduced separately before combining them into one map. 
The SPIRE data were largely reduced according to the standard pipeline (POF5\_pipeline.py, dated 8 Jun 2010), provided by the SPIRE Instrument Control Centre (ICC).  
Divergent from the standard pipeline were the use of the {\it sigmaKappaDeglitcher} (instead of the ICC-default 
{\it waveletDeglitcher}) and the BriGAdE method (Smith et al.\,in prep.)
to remove the temperature drift and bring all bolometers to the same level (instead of the default {\it temperatureDriftCorrection} and the residual,
median baseline subtraction).
Reduced SPIRE maps have pixel sizes of 6$\arcsec$, 8$\arcsec$ and 12$\arcsec$ at 250, 350 and 500 $\mu$m, respectively. The FWHM of the SPIRE beams are 18.2$\arcsec$, 24.5$\arcsec$ and 36.0$\arcsec$ at 250, 350 and 500 $\mu$m, respectively. SPIRE images are converted from MJy/beam to Jy/pixel units assuming beam areas of 423, 751 and 1587 arcsec$^2$ at 250, 350 and 500 $\mu$m, respectively. 
Since the calibration in the standard pipeline is optimized for point sources, we apply correction factors (0.9828, 0.9834 and 0.9710 at 250, 350 and 500 $\mu$m) to convert the K4 colour correction factors from point source to extended source calibration \citep{SPIRE}.
Additionally, multiplicative colour correction factors (0.9924, 0.9991 and 1.0249 at 250, 350 and 500 $\mu$m) were applied and a correction factor of 1.0067 was used to update the fluxes in the 350 $\mu$m image to the latest v7 calibration product \citep{SPIRE}. SPIRE images were also background subtracted, in a similar way as for the PACS images.

\subsection{PACS spectroscopy}
PACS spectroscopy maps, [C{\sc{ii}}] 157.74 $\mu$m and [O{\sc{i}}] 63 $\mu$m, were observed on the 14th of February 2011 (ObsID 1342214374, 1342214375, 1342214376). The [C{\sc{ii}}] observations cover the northern and the central area of NGC\,205 (see Figure \ref{specmap}), while two smaller [O{\sc{i}}] maps are centered on the CO peak in the north and on the centre of NGC\,205 (see Figure \ref{specmap}).

The PACS \citep{2010A&A...518L...2P} spectroscopic observations of NGC205 were done in the chop/nod mode, and cover an area of $95 \times 95$ arcsec ($3 \times 3$ pointings) and $47 \times 47$ arcsec (1 pointing) for the [C{\sc{ii}}] (157.74 $\mu$m) and both [O{\sc{i}}] (63 $\mu$m) maps, respectively. We used the largest chop throw of 6~arcmin to ensure we were not chopping onto extended source emission.  The maps were processed from Level 0 to Level 2 using the standard pipeline in HIPE (version 7.0.0), with version (FM, 32) of the calibration files.  Once the data are processed to Level2, we used the PACSman program\footnote{PACSman is available for download at http://www.myravian.fr/Homepage/Softwares.html.} \citep{Vianney} to perform line fits to the unbinned spectral data in each spatial pixel using a least-squares fitting routine, and to create integrated flux density maps of the results.  For [C{\sc{ii}}], the flux map mosaic was created by projecting the individual rasters onto an oversampled grid. More details can be found in \citet{Vianney}.

\setcounter{figure}{0}
\begin{figure}
\includegraphics[width=85mm,angle=0]{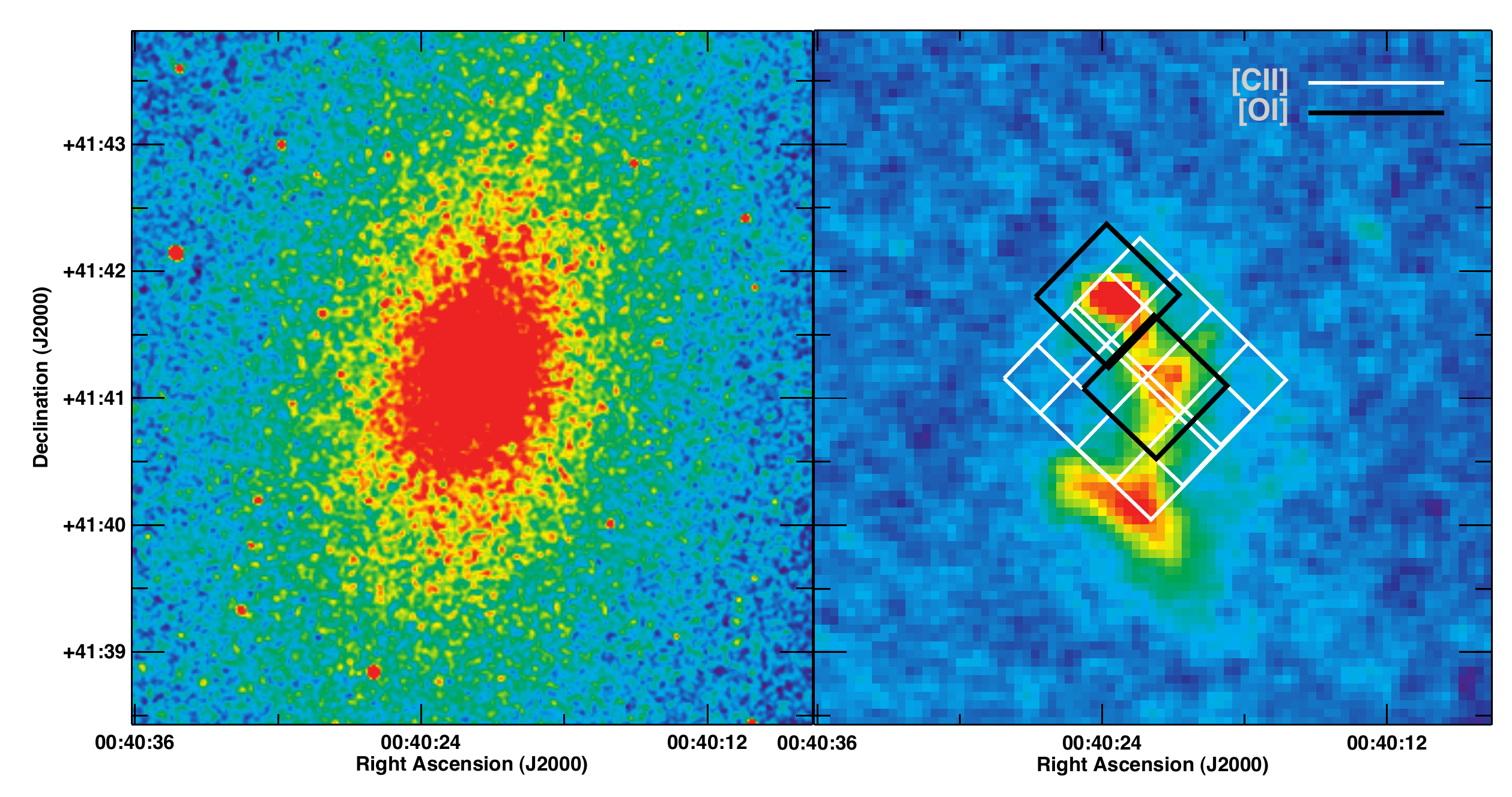} 
\caption{Left panel: $K$ band image of NGC\,205. Right panel: PACS 160 $\mu$m image overlaid with the AORs for the [O{\sc{i}}] (black, solid lineboxes) and [C{\sc{ii}}] (raster with white, solid lines) PACS line spectroscopy observations. The [O{\sc{i}}] line was observed in the north of NGC\,205, where the CO(1-0) emission peaks in the north, and in the central region of NGC\,205. The [C{\sc{ii}}] observations cover both areas.}
\label{specmap}
\end{figure}

\subsection{Noise calculations}
\label{Error.sec}
To determine the uncertainty on the PACS and SPIRE photometry data, we need to take into account three independent noise measurements. Aside from the most important uncertainty factor due to the calibration, the random background noise and the map making uncertainty contribute to the flux uncertainty as well.
For the calibration uncertainty, values of 3, 3, and 5 $\%$ were considered for the PACS 70, 100 and 160 $\mu$m data, respectively (PACS Observer's Manual 2011). All SPIRE wavebands were assumed to have a calibration uncertainty of 7 $\%$ (\citealt{2010A&A...518L...4S}, SPIRE Observer's Manual 2010). 
An estimate for the background noise is derived by taking 100 random apertures in the field around the galaxy (i.e. the same apertures used to determine the background), calculating the mean pixel value within each aperture as well the average mean over all apertures and, finally, computing the standard deviation of the mean pixel values in those individual apertures.
Map making uncertainties are derived from the error map, which is produced during the data reduction procedure. Specific uncertainties in map making are determined from this error map. The total uncertainty on the flux value in a pixel is calculated as the square root of the sum of the three squared error contributions. Calibrations and background uncertainties are summarized in Table \ref{fluxes}.

The noise level in PACS spectroscopy observations is determined from the uncertainty in the integrated intensity map, calculated from the the formula:
\begin{equation}
\Delta I = \left[ \Delta v \sigma \sqrt{N_{line}} \right] \sqrt{1+\frac{N_{line}}{N_{base}}}
\end{equation}
where $\Delta v$ corresponds to the channel width in km s$^{-1}$, $\sigma$ is the uncertainty in K and $N_{line}$ and $N_{base}$ represent the total number of channels covering the spectral line and the channels used for the baseline fitting, respectively.
A 1$\sigma$ uncertainty value is estimated from taking the mean over 15 random apertures in this uncertainty map (see Table \ref{noisespec}).

\begin{table}
\caption{The 1$\sigma$ noise levels for the PACS spectroscopy observations.}
\label{noisespec}
\begin{tabular}{@{}|lcc|}
\hline 
Line & Rest wavelength & $\sigma$ \\
\hline \hline
& $\mu$m & (10$^{-6}$ erg s$^{-1}$ cm$^{-2}$ sr$^{-1}$)  \\
\hline 
[C{\sc{ii}}] & 157.74 & 1.28 \\

[O{\sc{i}}] (center) & 63.18 & 5.50 \\

[O{\sc{i}}] (north) & 63.18 & 7.06 \\
\hline
\end{tabular}
\newline
\end{table}

\subsection{JCMT observations and ancillary data}
The JCMT observations of the $^{12}$CO~(3-2) transition (rest frequency 345.79 GHz) were obtained with the HARP-B instrument \citep{2009MNRAS.399.1026B} as part of project M10AC07 (PI: Tara Parkin) over eight nights in May, June and September of 2010, with a telescope beam size of 14.5~arcsec.  We obtained a single map in jiggle-chop mode with a footprint of 2~arcmin~$\times$~2~arcmin on the sky, and a total integration time of 1350~sec for each of the 16 jiggle positions.  The observations were carried out using beam-switching with a chop throw of 150~arcsec from the centre of NGC\,205.  We used the Auto-Correlation Spectrometer Imaging System (ACSIS) as our backend receiver, and it was set to a bandwidth of 1~GHz with 2048 channels, resulting in a resolution of 0.43 km~s$^{-1}$.  The data were then reduced using the Starlink\footnote{The S\textsc{tarlink} package is available for download at http://starlink.jach.hawaii.edu.} software package \citep{2008ASPC..394..650C}, maintained by the Joint Astronomy Centre.  For a full description of our data reduction and map making methods see \citet{2010ApJ...714..571W} and \citet{parkin_2011_submit}.

We obtained raw MIPS data for NGC\,205 from the \textit{Spitzer} archive, which were reprocessed according to the procedure outlined in \citet{2012arXiv1202.4629B}.
An H{\sc{i}} map for NGC\,205 obtained from VLA observations \citep{1997ApJ...476..127Y} was kindly provided to us by Lisa Young.

\section{Results}
\label{Results.sec}   

\subsection{Distribution of gas and dust in the ISM of NGC\,205}
\label{Distribution.sec}   
Figure \ref{vngsmaps} displays the \textit{Herschel} maps for NGC\,205 in the PACS 70, 100, 160 $\mu$m and SPIRE 250, 350 and 500 $\mu$m wavebands. 
In all bands, we are able to distinguish three dominant emission regions (north, central and south), which were first identified using MIPS data by \citet{2006ApJ...646..929M}. A substantial amount of dust also resides in between those three distinctive emission regions. 
Towards the southeast of NGC\,205 there is also an indication for a tentative detection of a colder dust component in the SPIRE maps (see the red, dashed ellipse in Figure \ref{vngsmaps}). The detection is below the 3$\sigma$ level, but it coincides with an optical tidal tail reported in \citet{2010IAUS..262..426S}.
However, we cannot rule out the possibility that the faint blob corresponds to foreground Galactic cirrus emission similar to the emission identified in the surroundings of M81 \citep{2010A&A...516A..83S,2010MNRAS.409..102D} or to emission originating from one or multiple background sources.

 \setcounter{figure}{1}
\begin{figure*}
\includegraphics[width=180mm,angle=0]{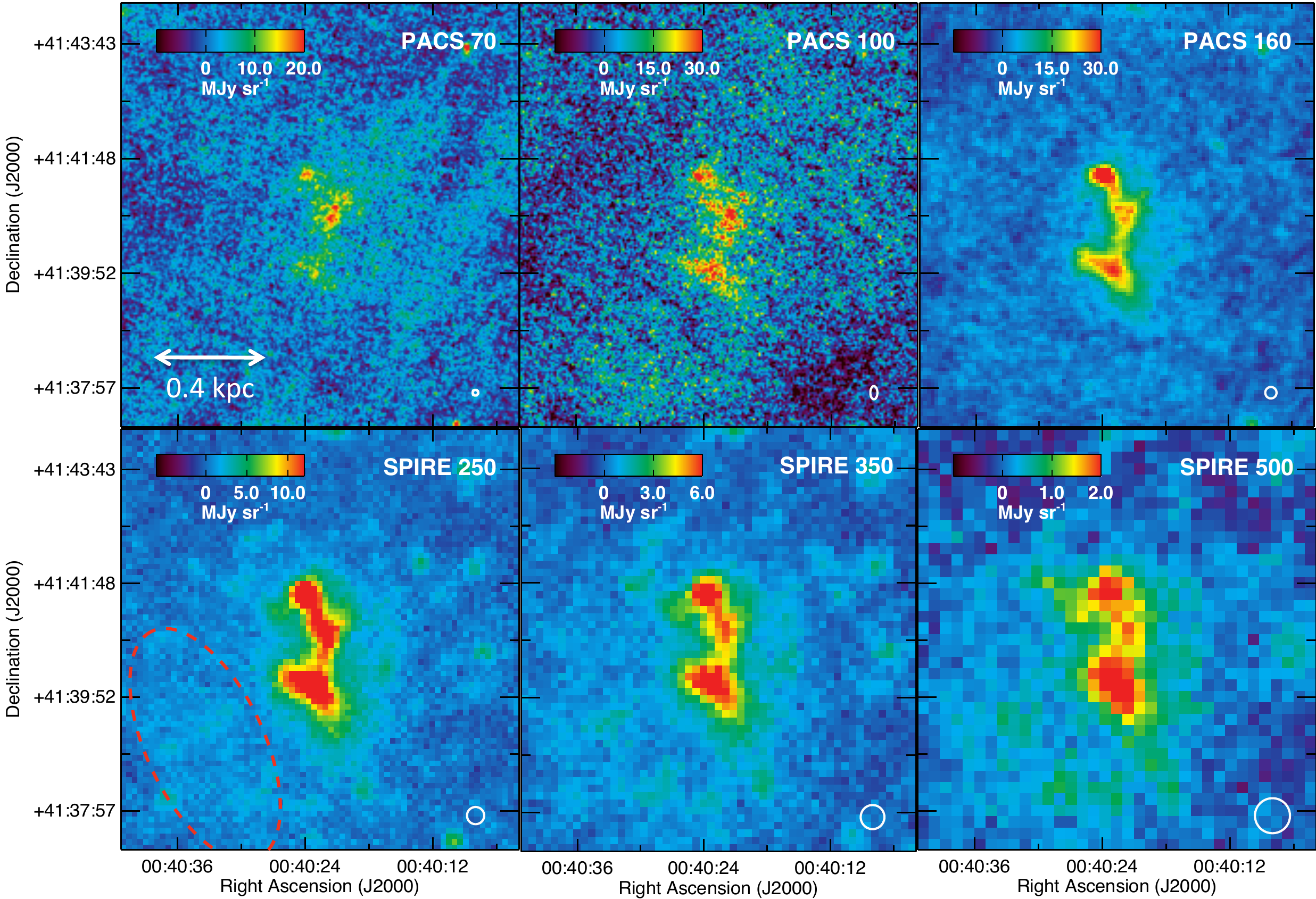} 
\caption{An overview of the Herschel maps. From left to right: PACS 70, 100, 160 $\mu$m (first row) and SPIRE 250, 350, 500 $\mu$m (second row). 
The displayed maps cover an area of 8.0$\arcmin$ $\times$ 6.5$\arcmin$, with north up and east to the left. Besides the PACS 70 $\mu$m (VNGS) and PACS 100 $\mu$m (HELGA) maps, all images in the other bands were produced using both VNGS and HELGA observations.
The FWHM of the PSF is indicated as a white circle in the lower left corner of each image. The red dashed line indicates the area where there is an indication for a tentative detection of a colder dust component.}
\label{vngsmaps}
\end{figure*}

When comparing the H{\sc{i}}, H$_{2}$ and dust distribution in NGC\,205 (see Figure \ref{hicontourmap}), we find a remarkable correspondence between the peaks in H{\sc{i}} and H$_{2}$ (as derived from the CO(3-2) observations, see Section \ref{Gas.sec}) column density and dust emission. This seems to imply that the dust component in NGC\,205 is well mixed with the atomic and molecular gas at the observed spatial scales of $\sim$ 100 pc for SPIRE. While a significant part of the atomic gas resides in the area south of the galaxy's center (see Figure \ref{hicontourmap}), current CO observations only cover most of the northern part of the galaxy (\citealt{1983ApJ...275..549J, 1997ApJ...476..127Y, 1998ApJ...499..209W}, JCMT CO(3-2) map from this work) and a minor part in the south of NGC\,205 \citep{1996ApJ...464L..59Y}. This lack of data makes it difficult to draw conclusions about the correlation of the molecular gas component with the dust or H{\sc{i}} gas in the southern area of NGC\,205.
At least for the northern part of NGC\,205, it seems that also the molecular gas component correlates well with the H{\sc{i}} gas and dust.

 \setcounter{figure}{2}
\begin{figure}
\includegraphics[width=85mm,angle=0]{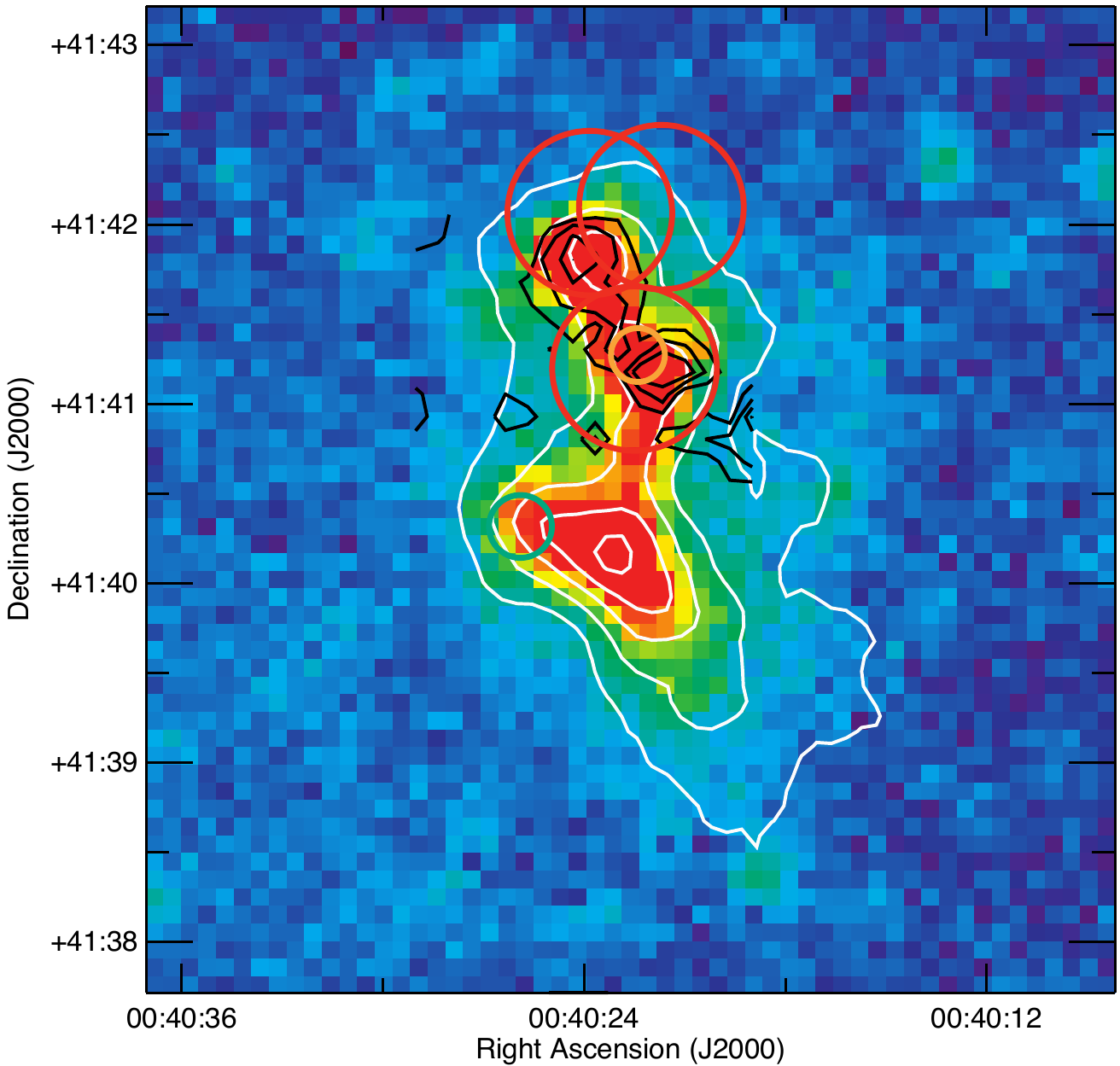} 
\caption{H{\sc{i}} column density contours (white, solid line, \citealt{1997ApJ...476..127Y}), H$_{2}$ column density contours (black, solid contours, derived from the JCMT CO(3-2) map), CO(1-0) pointings (red circles: \citealt{1998ApJ...499..209W}, green circle: \citealt{1996ApJ...464L..59Y}) and JCMT 1.1 mm 18$\arcsec$ beam (yellow circle, \citealt{1991ApJ...374L..17F}) overlaid on the SPIRE 250 $\mu$m image. The H{\sc{i}} contours range from 2 $\times$ 10$^{19}$ to 3.5 $\times$ 10$^{20}$ cm$^{-2}$ in intervals of 6.6 $\times$ 10$^{19}$ cm$^{-2}$, while the CO(3-2) contours represent a H$_{2}$ column density range 2.6 $\times$ 10$^{20}$ $\leq$ $N_{\text{H}_{2}}$ $\leq$ 1.9 $\times$ 10$^{21}$ cm$^{-2}$ increased in steps of 3.3 $\times$ 10$^{20}$ cm$^{-2}$.}
\label{hicontourmap}
\end{figure}

\subsection{Global fluxes}
\label{Global.sec}  
Global fluxes in all wavebands were determined from summing over all pixels in the background subtracted image with $>$ 3$\sigma$ detection in the SPIRE 250 $\mu$m image.
Those global fluxes and the corresponding noise measurements (calibration uncertainty and random background noise) in each waveband are summarized in Table \ref{fluxes}. Flux measurements have been updated to the last PACS and SPIRE calibration products, converted to the extended source calibration and color corrected for each filter. 

When comparing our fluxes at PACS 70 $\mu$m (2.2 $\pm$ 0.2 Jy) and PACS 160 $\mu$m (4.0 $\pm$ 0.3 Jy) to the MIPS fluxes at those overlapping wavelengths reported in \citet{2006ApJ...646..929M} (MIPS 70 $\mu$m: 1.4 $\pm$ 0.3 Jy; MIPS 160 $\mu$m: 8.8 $\pm$ 4.7 Jy), a large discrepancy between PACS and MIPS fluxes is found. In view of this large difference, we determined MIPS fluxes (MIPS 70 $\mu$m: 1.3 $\pm$ 0.3 Jy; MIPS 160 $\mu$m: 4.6 $\pm$ 0.9 Jy) from our reprocessed archival MIPS data by summing over the same pixels (uncertainty values only refer to calibration uncertainties). To determine the corresponding PACS fluxes, the Herschel maps at 70 and 160 $\mu$m wavelengths were convolved with the appropriate kernels to match the elongated wings of the MIPS PSFs. The customized kernels were created following the procedure in \citet{2011arXiv1109.0237B} (see also \citealt{2008ApJ...682..336G}). Summing the flux values for the same pixels, we find the corresponding flux measurements for PACS 70 $\mu$m (2.3 $\pm$ 0.2 Jy) and PACS 160 $\mu$m (3.8 $\pm$ 0.3 Jy). Upon comparison of the fluxes, we find a relatively good agreement between the PACS measurement and the flux determined from the archival MIPS image at 160 $\mu$m. The PACS 70 $\mu$m flux is sufficiently higher, which might either be a calibration issue or a dissimilarity in the background determination. 
We argue that the deviation from the MIPS 160 $\mu$m measurement reported in \citet{2006ApJ...646..929M} is either due to a flux calibration issue or an overestimated aperture correction.
Indeed, the flux calibration for MIPS \citep{2007PASP..119.1019G, 2007PASP..119.1038S} was only finalized after the analysis in \citet{2006ApJ...646..929M}.

A similar comparison at 100 $\mu$m is possible between PACS and IRAS fluxes. \citet{1988ApJS...68...91R} report a total flux density of 3.78 $\pm$ 0.57 Jy for NGC\,205, which agrees well with the flux density 3.6 $\pm$ 0.5 Jy determined from \textit{Herschel} observations.

\begin{table}
\caption{Global fluxes ($F_{\nu}$) and contributions from the random background noise ($\sigma_{back}$) and the calibration uncertainty ($\sigma_{cal}$) are provided for every waveband. All tabulated fluxes have been multiplied by extended source calibration and filter colour correction factors.}
\label{fluxes}
\begin{tabular}{@{}|lccc|}
\hline 
Wavelength & $F_{\nu}$  & $\sigma_{back}$ & $\sigma_{cal}$\\
\hline \hline
$\mu$m & (mJy) & ($\mu$Jy/$\arcsec^{2}$) & ($\%$) \\
\hline 
MIPS 24 & 98 $\pm$ 15 &  & 15\footnotemark[1] \\
PACS 70 & 2204 $\pm$ 179 &  0.5 & 3 \\ 
PACS 100 & 3565 $\pm$ 515 &  1.2 & 3 \\
PACS 160 & 3912 $\pm$ 159 &  2.2 & 5 \\
SPIRE 250 & 2623 $\pm$ 29 &  1.8 & 7 \\
SPIRE 350 & 1302 $\pm$ 15 & 1.6 & 7 \\
SPIRE 500 & 539 $\pm$ 9 &  2.3 & 7 \\
\hline
\end{tabular}

\footnotemark[1] {The 15$\%$ uncertainty on the MIPS 24 $\mu$m flux includes the uncertainties arising from calibration and background subtraction.}
\end{table}

\subsection{Calculation of the dust mass}
\label{Dust.sec}  
\subsubsection{SED fitting method}
\label{SEDfitting.sec}   
For the SED fitting procedure, we apply the DustEm code \citep{2011A&A...525A.103C}, which predicts the emission of dust grains given the strength of the interstellar radiation field (ISRF) and a certain composition of grain types, with a specific size distribution, optical and thermal dust properties. 
DustEm derives the local dust emissivity from computing explicitly the temperature distribution for every grain type of particular size and composition. For the analysis in this paper, we adopt two different dust compositions (\citealt{2007ApJ...657..810D} and \citealt{2011A&A...525A.103C}), each containing polycyclic aromatic hydrocarbons (PAHs) and amorphous silicate grains complemented with either graphite or amorphous carbon dust particles, respectively. The dust composition from \citet{2007ApJ...657..810D} corresponds to the typical dust mixture found in our own Galaxy. The amorphous carbonaceous grains are in the form of hydrogenated amorphous carbon, better known as a-C:H or HAC \citep{2011A&A...525A.103C}. The spectral shape of the ISRF is assumed to be the same as determined in the solar neighborhood \citep{1983A&A...128..212M} (MMP83). 
Although the shape and hardness of the ISRF in low-metallicity dwarf galaxies might differ from the Galactic ISRF \citep{2006A&A...446..877M}, altering the spectral shape of the ISRF will in particular influence the radiation of transiently heated PAHs and very small grains but has been shown to only affect the total dust mass estimate by a factor of $<$ 10$\%$ \citep{2009A&A...508..645G}.  
Moreover, based on the number and spectral type of young massive stars in NGC\,205, \citet{1996ApJ...464L..59Y} found an UV field corresponding well to the ISRF in the solar neighborhood.
Under these assumptions, we have only 2 free parameters: the dust mass $M_{\text{d}}$ and the intensity of the ISRF $X_{\text{ISRF}}$ relative to the Galactic ISRF.

During the SED fitting procedure, we explore a parameter grid in $X_{\text{ISRF}}$ and dust mass by increasing them stepwise by a factor of 1.05. In order to facilitate the least-square fitting procedure, we construct a pre-calculated library of dust models in DustEm, each with a different scaling of the ISRF and dust mass.
The model SED at those wavelengths is convolved with the response function of the filter-band passes, to include the appropriate colour correction (pixel values were not colour corrected in this case). Finally, the model with the best fitting parameters is determined from a least-square fitting routine.
To estimate the uncertainties on the best fitting parameters, we perform a bootstrapping procedure. Hereto, the same fitting routine is applied on a dataset of 100 flux densities, randomly determined from a Gaussian distribution for which the maximum value and width are chosen to correspond to the observed fluxes and uncertainty values. The 1$\sigma$ uncertainties on our best fitting model parameters correspond to the 16 and 84 percentiles of this Gaussian distribution.

Since the primary goal of this analysis is quantifying the total dust content in NGC\,205, we restrict our SED fitting procedure to the wavelength range from 24 to 500 $\mu$m. The spectral shape throughout this wavelength range is determined from the flux measurements in the six available \textit{Herschel} wavebands (PACS 70, 100, 160 $\mu$m and SPIRE 250, 350, 500 $\mu$m) and the MIPS 24 $\mu$m flux. 
Prior to SED fitting, all images (Herschel+ancillary MIPS data) were convolved with the appropriate kernels, according to the procedure in \citet{2011arXiv1109.0237B}, to match the resolution of the 500 $\mu$m images. All images were also rebinned to the pixel scale (12$\arcsec$/pixel) of the 500 $\mu$m image. 

\subsubsection{Pixel-by-pixel fitting}
\label{pixelbypixel.sec}
Dust masses and ISRF scaling factors are computed from a SED fitting procedure to every pixel with fluxes in at least three different bands above the 3$\sigma$ level to constrain the spectral shape of the energy distribution. In this way, we obtain 155 pixels with a sufficient signal to noise level and avoid contribution from noisy pixels (in particular at 100 $\mu$m) to the SED fitting procedure, which might bias our estimate of the total dust mass. The uncertainty on the flux values in every pixel are calculated following the procedure outlined in Section \ref{Error.sec}.

\setcounter{figure}{3}
\begin{figure}
\includegraphics[width=85mm,angle=0]{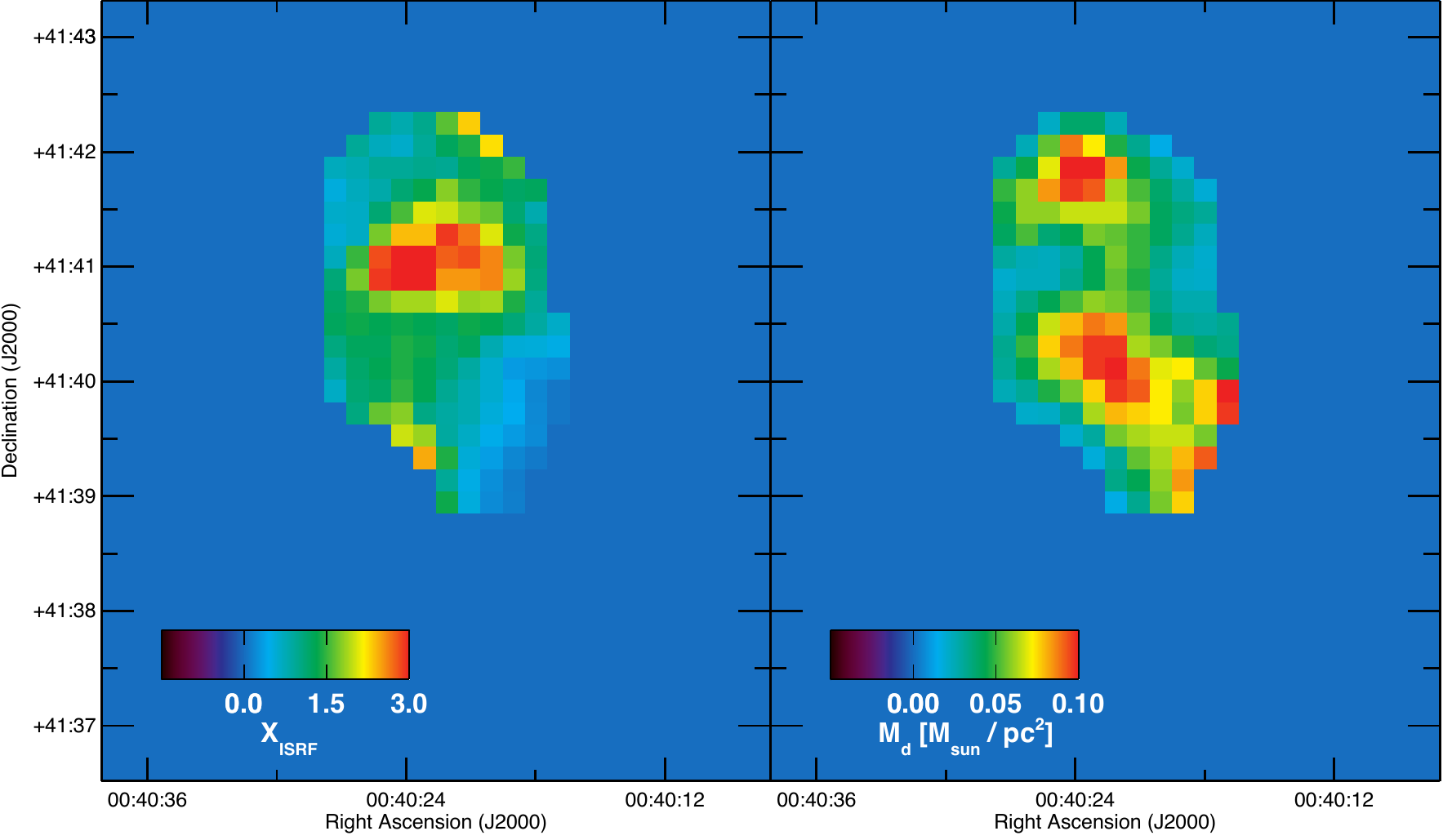} 
\caption{The $X_{\text{ISRF}}$ (left panel) and dust mass (right panel) obtained from a multi-component DustEm SED fitting procedure for every pixel with at least 3 detections above the 3$\sigma$ level across the 24 $\mu$m to 500 $\mu$m wavelength range. }
\label{pixelbypixel}
\end{figure}

Figure \ref{pixelbypixel} shows the maps with the best fitting ISRF scaling factors (left) and dust column densities (right), determined from SED fitting with the silicate+amorphous carbon dust composition \citep{2011A&A...525A.103C}.
From those maps, we clearly notice that the dust in the center of NGC\,205 has a higher temperature, while most of the dust mass resides in the northern and southern regions of NGC\,205, corresponding to peaks in the H{\sc{i}}+H$_{2}$ and H{\sc{i}} column density, respectively. 
For the three distinctive dust emission regions in NGC\,205 the northern and southern areas are brighter because of the large amounts of dust residing in these areas, while the dust grains in the central region emit more prominently because they are exposed to a stronger ISRF. Also outside these three emission regions, a substantial amount of dust is found to be present in the galaxy. 

The total dust mass in NGC\,205 is determined by adding the dust masses in all pixels, resulting in $M_{\text{d}}$ $\sim$ 1.8 $\times$ 10$^{4}$ M$_{\odot}$.
An estimate for the average strength of the ISRF is calculated as the median of the values for $X_{\text{ISRF}}$ in all pixels. This median value of $X_{\text{ISRF}}$ $\sim$ 1.13 is translated into a large grain dust temperature $T_{d}$ $\sim$ 17.8 K by averaging over all mean temperatures for grains with sizes between 4 $\times$ 10$^{-3}$ and 2 $\mu$m for this specific strength of the ISRF. 
When repeating the SED fitting procedure with a silicate+graphite dust composition \citep{2007ApJ...657..810D}, we find values for the dust mass ($M_{\text{d}}$ $\sim$ 1.1 $\times$ 10$^{4}$ M$_{\odot}$) and the median value of the scaling factor $X_{\text{ISRF}}$ $\sim$ 3.12 (or $T_{d}$ $\sim$ 21.2 K) consistent with the results obtained for a silicate+amorphous carbon dust composition within the uncertainties of the SED fitting procedure.
The fitting results for both dust compositions are summarized in Table \ref{parmodels}.
To check whether our results are hampered by the resolution in the SPIRE 500 $\mu$m waveband ($\sim$ 36$\arcsec$), we perform the same pixel-by-pixel SED fitting procedure at a resolution of the SPIRE 350 $\mu$m waveband ($\sim$ 24.5 $\arcsec$). Whereas SED fitting on higher resolution data have shown to probe a more massive dust reservoir in the Large and Small Magellanic Clouds \citep{2011arXiv1110.1260G}, we find best fitting values for the average scaling factor $X_{\text{ISRF}}$ $\sim$ 0.61$^{+0.31}_{-0.31}$ and the total dust mass $M_{d}$ $\sim$ 0.9$^{+0.5}_{-0.5}$ $\times$ 10$^{4}$ M$_{\odot}$. Since the dust mass is somewhat lower than the results obtained from the SED fit with the 500 $\mu$m measurement, we conclude that a gain in resolution does not better trace the dust content in NGC\,205. On the contrary, including the 500 $\mu$m data in the SED fit results in a more massive dust component at a colder temperature in the outer regions of NGC\,205.

A SED fitting procedure for the global fluxes was performed as well, giving similar results as for the pixel-by-pixel dust masses and $X_{\text{ISRF}}$ factors (see Table \ref{parmodels}). Figure \ref{sed_bright2} displays the best fitting DustEm model for the silicate+amorphous carbon dust composition overlaid
with the global \textit{Herschel} fluxes and other data from the literature. 

\subsubsection{Submm/mm excess}
The dust masses obtained from our \textit{Herschel} observations ($M_{\text{d}}$ $\sim$ 1.1-1.8 $\times$ 10$^{4}$ M$_{\odot}$) are comparable to the dust masses derived from the MIPS observations ($M_{\text{d}}$ $\sim$ 3 $\times$ 10$^{4}$ M$_{\odot}$) within the uncertainties of the observations and fitting procedure, but more than one order of magnitude lower than the predicted dust mass ($M_{\text{d}}$ $\sim$ 5 $\times$ 10$^{5}$ M$_{\odot}$) at a temperature of $\sim$ 12 K based on mm+\textit{Spitzer} observations \citep{2006ApJ...646..929M}. 
Since the strength of the ISRF and therefore the heating of the dust grains is variable throughout the plane of the galaxy (see Figure \ref{pixelbypixel}, left panel), we calculate an upper limit for the dust mass at a dust temperature of $T_{d}$ $\sim$ 12 K by scaling the SED until fitting the upper limit of the 500 $\mu$m flux measurement ($F_{\nu}$ $\sim$ 612 mJy). We derive an upper mass limit $M_{d}$ $<$ 4.9 $\times$ 10$^{4}$ M$_{\odot}$ for the cold dust reservoir ($T_{d}$ $\sim$ 12 K). Therefore, our \textit{Herschel} observations do not seem to support the presence of a massive cold dust component, implying that the millimeter flux is unlikely to originate from a cold dust reservoir.  

With an upper limit of 0.06 mJy at 21 cm \citep{2007AJ....134.2148L} and the absence of emission lines characteristic for LINER or Seyfert galaxies \citep{2011arXiv1110.5891M}, the JCMT 1.1 mm flux measured by \citet{1991ApJ...374L..17F} is unlikely to be caused by synchrotron emission from either supernova remnants or an AGN-like nucleus. Also the contribution from CO(2-1) line emission seems unable to account for the high mm measurement.
Indeed, based on the JCMT CO(2-1) line intensity (0.43 K km/s) for the inner region \citep{1998ApJ...499..209W}, we derive a flux density of 0.724 Jy at 1.3 mm. Relying on the narrow CO(2-1) line width (13 km/s or 10 MHz) and the bandwidth of the IRAM UKT14 receiver (74 GHz), a contribution from CO(2-1) line emission to the 1.1 mm continuum observations is found to be negligible. This implies that other explanations (calibration issues, background source, bad weather conditions) need to be invoked to explain the high 1.1 mm flux in the center of NGC\,205.

Also \textit{Herschel} observations at 500 $\mu$m do not show any indication for excess emission at submm or mm wavelengths, reminiscent of the submm excess observed in many star-forming dwarf galaxies and blue compact dwarfs \citep{2003A&A...407..159G,2004A&A...414..475D,2009A&A...508..645G,2010A&A...518L..55G,2011A&A...532A..56G,2010A&A...518L..52G, 2010A&A...518L..58O}.  
In some cases, the observed excess even extends up to millimeter and centimetre wavelengths, such as observed in the Large (LMC) and Small Magellanic Clouds (SMC) \citep{2010A&A...519A..67I,2010A&A...523A..20B,2010A&A...518L..71M,2011arXiv1101.2046P}. Several reasons have been invoked to account for this excess emission in the submm/mm wavebands. Either large amounts of very cold dust (e.g. \citealt{2003A&A...407..159G, 2005A&A...434..867G, 2009A&A...508..645G, 2010A&A...518L..58O}), dust grains with optical properties diverging from the typical Galactic dust characteristics (e.g. \citealt{2002A&A...382..860L,2007A&A...468..171M}) or spinning dust grains (e.g. \citealt{2010A&A...523A..20B,2011arXiv1101.2046P}) are thought to be responsible for the excess emission.
The fact that our SED model can account for the observed 500 $\mu$m emission in NGC\,205 is interesting, because it is in contrast with the submillimeter excess observed in several other low metallicity dwarf galaxies.
This might be an indication for different ISM properties and star-forming conditions in this early-type dwarf galaxy, compared to the typical star-forming dwarfs revealing an excess submm emission. However, an excess in wavebands longwards of 500 $\mu$m (see also \citealt{2011A&A...532A..56G}) cannot be ruled out based on the currently available dataset for NGC\,205.

\begin{table}
\caption{Overview of the parameters (scaling factor for the ISRF $X_{\text{ISRF}}$, dust mass $M_{\text{d}}$) for the best fitting DustEm model, either with a silicate+amorphous carbon or silicate+graphite dust composition, determined from a pixel-by-pixel or global SED fit. The temperature estimate is obtained by taking the average over all mean temperatures for dust particles with sizes between 4 $\times$ 10$^{-3}$ and 2 $\mu$m.}
\label{parmodels}
\begin{tabular}{@{}|lccc|}
\hline 
Silicate/Am. Carbon &  $X_{\text{ISRF}}$ & $T_{d}$ [K] & $M_{\text{d}}$ [M$_{\odot}$]\\
\hline \hline 
Pixel-by-pixel & 1.13$^{+0.07}_{-0.07}$  & 17.8$^{+0.2}_{-0.1}$  & 1.8$^{+0.1}_{-0.1}$ $\times$ 10$^{4}$  \\
Global & 2.74$^{+0.42}_{-0.58}$  & 20.7$^{+0.5}_{-0.8}$  & 1.1$^{+0.1}_{-0.1}$ $\times$ 10$^{4}$  \\
\hline \hline
Silicate/graphite & $X_{\text{ISRF}}$ & $T_{d}$ [K] & $M_{\text{d}}$ [M$_{\odot}$]\\
\hline \hline 
Pixel-by-pixel & 3.12$^{+0.19}_{-0.19}$ & 21.2$^{+0.2}_{-0.3}$  & 1.1$^{+0.1}_{-0.1}$ $\times$ 10$^{4}$  \\
Global & 3.74$^{+0.10}_{-1.37}$  & 21.8$^{+0.1}_{-1.6}$  & 1.0$^{+0.3}_{-0.1}$ $\times$ 10$^{4}$  \\
\hline
\end{tabular}
\end{table}

\setcounter{figure}{4}
\begin{figure*}
\includegraphics[width=170mm,angle=0]{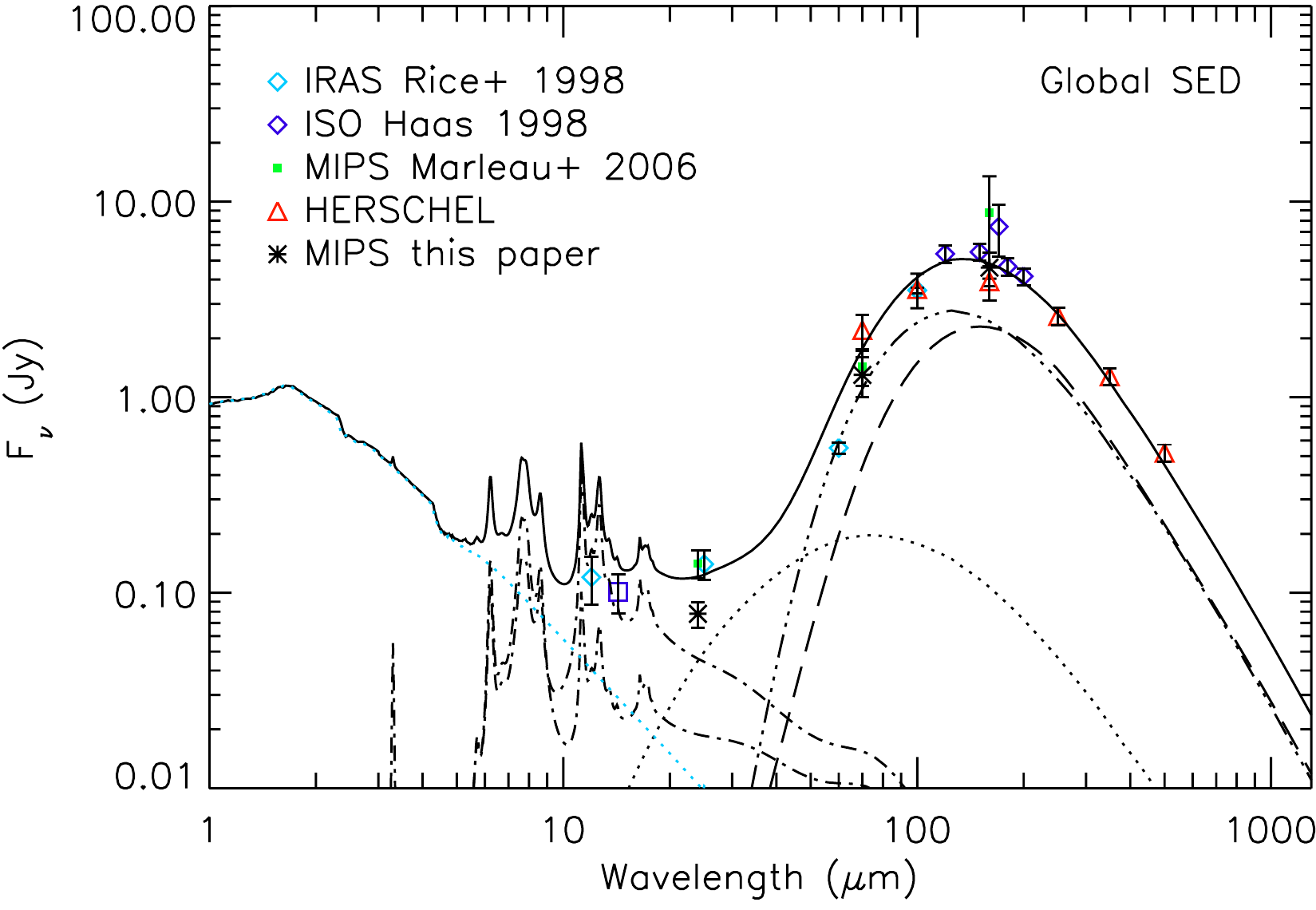} 
\caption{The SED for the best fitting DustEm model ($X_{\text{ISRF}}$ = 2.74 or $T_{d}$ $\sim$ 20.7 K, $M_{\text{d}}$ $\sim$ 1.1 $\times$ 10$^4$ M$_{\odot}$) with an silicate+amorphous carbon dust composition (black, solid line), overlaid with the measured MIPS 24, PACS 70, 100, 160 $\mu$m and SPIRE 250, 350 and 500 $\mu$m flux densities. Also the other flux densities from the literature are indicated.
The dotted-dashed, dotted, triple dotted-dashed and dashed lines correspond to the PaH dust mixtures, small and large amorphous carbon and silicate dust grains, respectively.
Since the SED fitting procedure with the silicate+graphite dust composition gave very similar results and uncertainties, the plot with this SED model is not explicitly shown here. Also a stellar component for NGC\,205, parametrized as a \citet{1998MNRAS.300..872M, 2005MNRAS.362..799M} Single Stellar Population with an age of 13 Gyr and a metallicity of Z = 0.002 \citep{1995ApJ...438..680B} and modeled as a Sersic profile of index n = 1 and effective radius 130$^{\prime \prime}$ \citep{2006MNRAS.369.1321D}, was included on the SED fit, to allow a comparison of our SED model to observations at NIR/MIR wavelengths (blue dotted line).
}
\label{sed_bright2}
\end{figure*}

\subsection{Gas mass}
\label{Gas.sec}   
\subsubsection{JCMT CO(3-2) observations}

In view of the limited coverage of previous CO(1-0) and CO(2-1) observations with only few pointings across the galaxy, we derive a molecular gas mass estimate for NGC\,205 from a CO(3-2) map covering a larger part of the galaxy, however still only probing the H$_{2}$ gas in the northern region of the galaxy (see Figure \ref{hicontourmap}).

For all pixels with detections $>$ 3$\sigma$, the integrated CO(3-2) line intensity is converted to a H$_{2}$ column density according to the formula:
\begin{equation}
N_{\text{H}_{2}} = \frac{X_{\text{CO}} I_{\text{CO(3-2)}}}{\eta_{\text{mb}}\left( \frac{I_{\text{CO(3-2)}}}{I_{\text{CO(1-0)}}} \right)}
\end{equation}
where $I_{\text{CO(3-2)}}$ is the total integrated line intensity expressed in units of K km s$^{-1}$. 
The scaling factor to convert an antenna temperature T$_{A}^{\star}$ into a main beam temperature $T_{mb}$ at the JCMT is $\eta_{mb}$ = 0.6. 
We assume a value of $\sim$ 0.3 for the CO(3-2)-to-CO(1-0) line intensity ratio corresponding to the typical ratios found in the diffuse ISM of other nearby galaxies \citep{2009ApJ...693.1736W}. Since we derive a line ratio of $\sim$ 0.34 for the central pointing reported in \citet{1998ApJ...499..209W}, we argue that this ratio serves as a good approximation for the entire CO(3-2) emitting region. The same $X_{\text{CO}}$ conversion factor (6.6 $\times$ 10$^{20}$ cm$^{-2}$ (K km s$^{-1}$)$^{-1}$) as introduced in Section \ref{gasobs.sec} is applied here.
The total molecular gas mass is derived from the column density following the equation:
\begin{equation}
\label{eqmco}
M_{\text{H}_{2}} = A N_{\text{H}_{2}} m_{\text{H}_{2}}
\end{equation}
where $A$ represents the surface of the CO(3-2) emitting region and $m_{\text{H}_{2}}$ is the mass of a molecular hydrogen atom.  
Inserting the correct values in equation \ref{eqmco} results in an estimate for the total molecular gas mass $M_{\text{H}_{2}}$ $\sim$ 1.3 $\times$ 10$^5$ M$_{\odot}$. Considering that this value is a factor of $\sim$ 5 lower than the $M_{\text{H}_{2}}$ $\sim$ 6.9 $\times$ 10$^5$ M$_{\odot}$ inferred from CO(1-0) detections, we argue that most of the H$_{2}$ in NGC\,205 resides in regions of colder temperature ($T$ $\leq$ $T_{\text{crit,CO(3-2)}}$ $\sim$ 33 K) and/or lower density ($n$ $\leq$ $n_{\text{crit,CO(3-2)}}$ $\sim$ 2 $\times$ 10$^4$ cm$^{-3}$). 

\subsubsection{PACS spectroscopy observations}

Relying on the low metal abundance of the ISM in the inner regions of NGC\,205, there might be a significant fraction of molecular gas in NGC\,205 which remains undetected by current CO observations, since CO is often a poor diagnostic of the H$_{2}$ content in low abundance environments exposed to hard radiation fields. Based on the high values for the $L_{\text{[CII]}}$-to-$L_{\text{CO}}$ ratio found in several low metallicity dwarf galaxies (e.g. \citealt{2000NewAR..44..249M,2010A&A...518L..57C}), [C{\sc{ii}}] is often claimed to be a better tracer for the molecular gas in such environments. Also the [O{\sc{i}}] fine-structure line, which is considered an important coolant of the neutral gas together with [C{\sc{ii}}], can be used as an alternative probe for the molecular gas in a low metallicity ISM (e.g. \citealt{1985ApJ...291..722T}).
Although both [C{\sc{ii}}] and [O{\sc{i}}] are considered good tracers of molecular gas in a metal-poor ISM, the interpretation of their line fluxes is hampered by a lack of knowledge about the exact origin of the line emission from within a galaxy.
[C{\sc{ii}}] emission is thought to arise either from the ionized (H{\sc{ii}} regions) or the neutral (PDRs) medium, while [O{\sc{i}}] emission originates mainly from the neutral ISM. In those neutral PDRs, the [C{\sc{ii}}] line provides cooling for gas clouds with a moderate density ($n_{\text{H}_{2}}$ $<$ 10$^4$ cm$^{-3}$), while the [O{\sc{i}}] line cools the higher density regions. 

\begin{table}
\caption{[C{\sc{ii}}] line measurements within the elliptical apertures overlaid on the [C{\sc{ii}}] map in Figure \ref{cii_map}}
\label{ciiflux}
\begin{tabular}{@{}|l|cc|cc|}
\hline 
Aperture & RA ($^{\circ}$) & DEC ($^{\circ}$) &  a ($\arcsec$) & b ($\arcsec$)  \\	         
 \hline
1   & 10.095534 & 41.673843 & 21.20 & 13.76 \\
2 & 10.090351 & 41.687053 & 33.84 & 23.45 \\
3 & 10.098251 & 41.69641 & 13.39 & 9.65 \\
                   \hline 
 \multicolumn{5}{c}{}          \\
                   \hline 
Aperture  & \multicolumn{2}{c|}{$F_{\nu}$ [10$^{-14}$ erg s$^{-1}$ cm$^{-2}$]} &  \multicolumn{2}{c|}{$M_{\text{H}_{2}}$ [M$_{\odot}$]}  \\
  \hline
1 & \multicolumn{2}{c|}{1.06 $\pm$ 0.25} & \multicolumn{2}{c|}{1.26 $\times$ 10$^3$} \\
2 & \multicolumn{2}{c|}{9.91 $\pm$ 0.52} & \multicolumn{2}{c|}{11.83 $\times$ 10$^3$} \\
3 & \multicolumn{2}{c|}{3.06 $\pm$ 0.56} & \multicolumn{2}{c|}{2.27 $\times$ 10$^3$} \\
\hline 
\end{tabular}
\newline
\end{table}

\setcounter{figure}{5}
\begin{figure*}
\includegraphics[width=89mm]{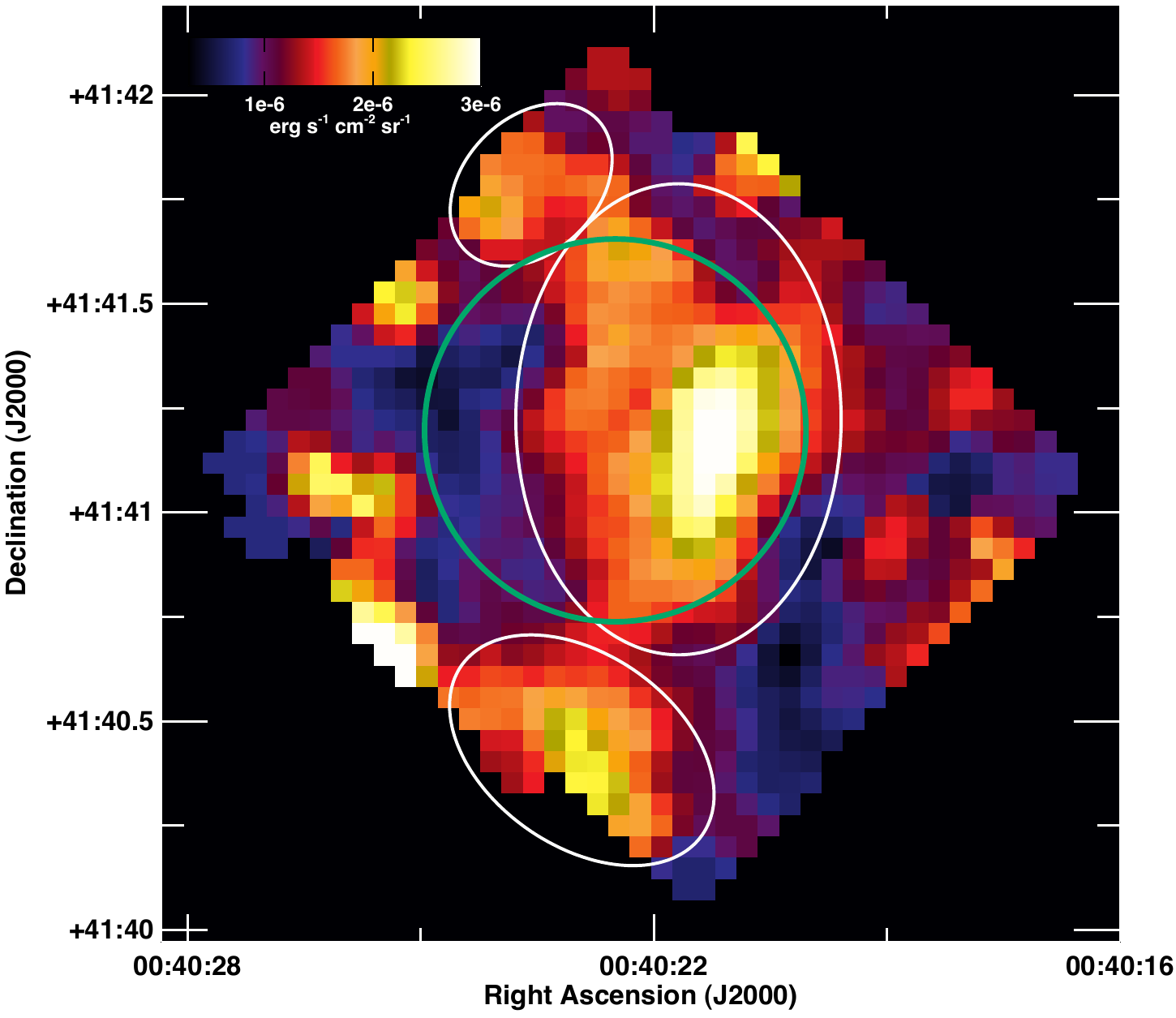} 
\includegraphics[width=85mm]{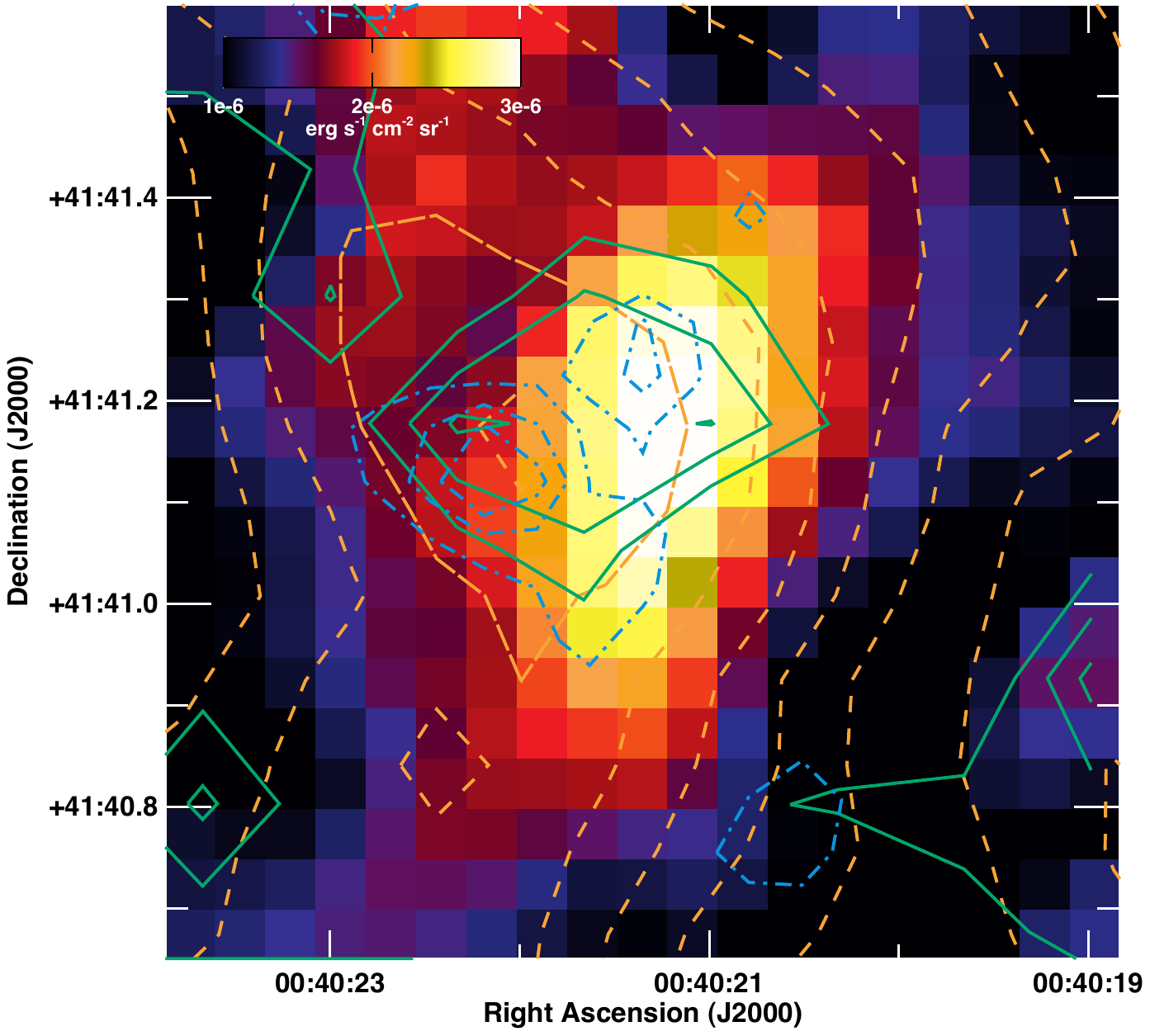} 
\caption{Left: PACS [C{\sc{ii}}] map of NGC\,205. The elliptical apertures for photometry are indicated as white circles. The CO(1-0) pointing reported in \citet{1998ApJ...499..209W} covering the center of NGC\,205 as well as the [C{\sc{ii}}] intensity peak is color-coded in green. Right: Zoom on the central bright region in the [C{\sc{ii}}] map, overlaid with contours of MIPS 24 $\mu$m surface brightnesses (blue, dashed-dotted curves) and H{\sc{i}} and CO(3-2) column densities (yellow, dashed and green, solid lines, respectively). The H{\sc{i}} contours range from 2 $\times$ 10$^{19}$ cm$^{-2}$ to 2.6 $\times$ 10$^{20}$ cm$^{-2}$ in intervals of 4 $\times$ 10$^{19}$ cm$^{-2}$, while the CO(3-2) contours represent a H$_{2}$ column density range 1.83 $\times$ 10$^{20}$ $\leq$ $N_{\text{H}_{2}}$ $\leq$ 1.28 $\times$ 10$^{21}$ cm$^{-2}$ increased in steps of 2.75 $\times$ 10$^{20}$ cm$^{-2}$. The contours representing the MIPS 24 $\mu$m surface brightnesses range from 0.68 to 2 MJy/sr, stepwise increased by 0.33 MJy/sr.}
\label{cii_map}
\end{figure*}

\setcounter{figure}{6}
\begin{figure*}
\includegraphics[width=35.5mm]{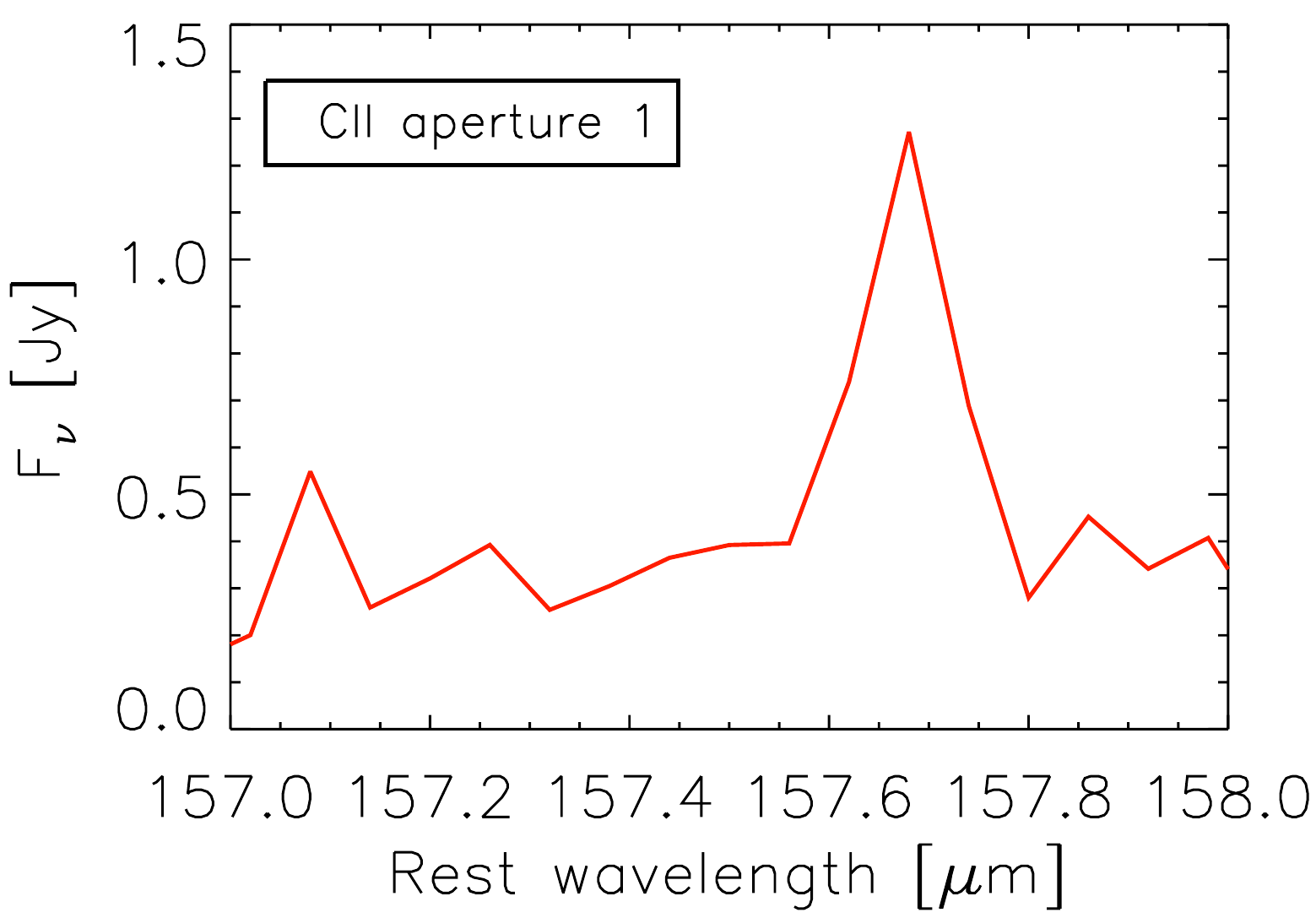} 
\includegraphics[width=34.0mm]{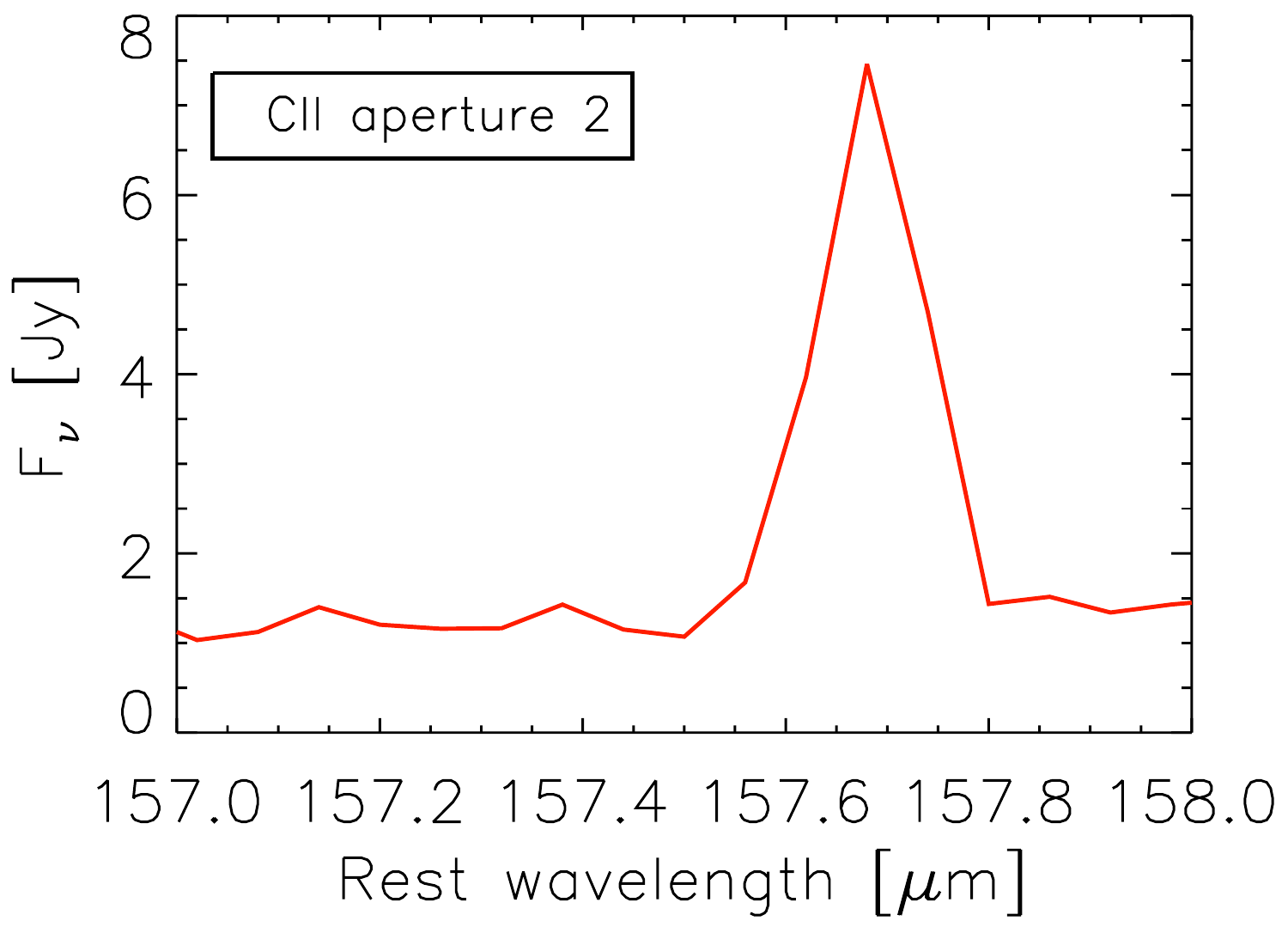} 
\includegraphics[width=34.0mm]{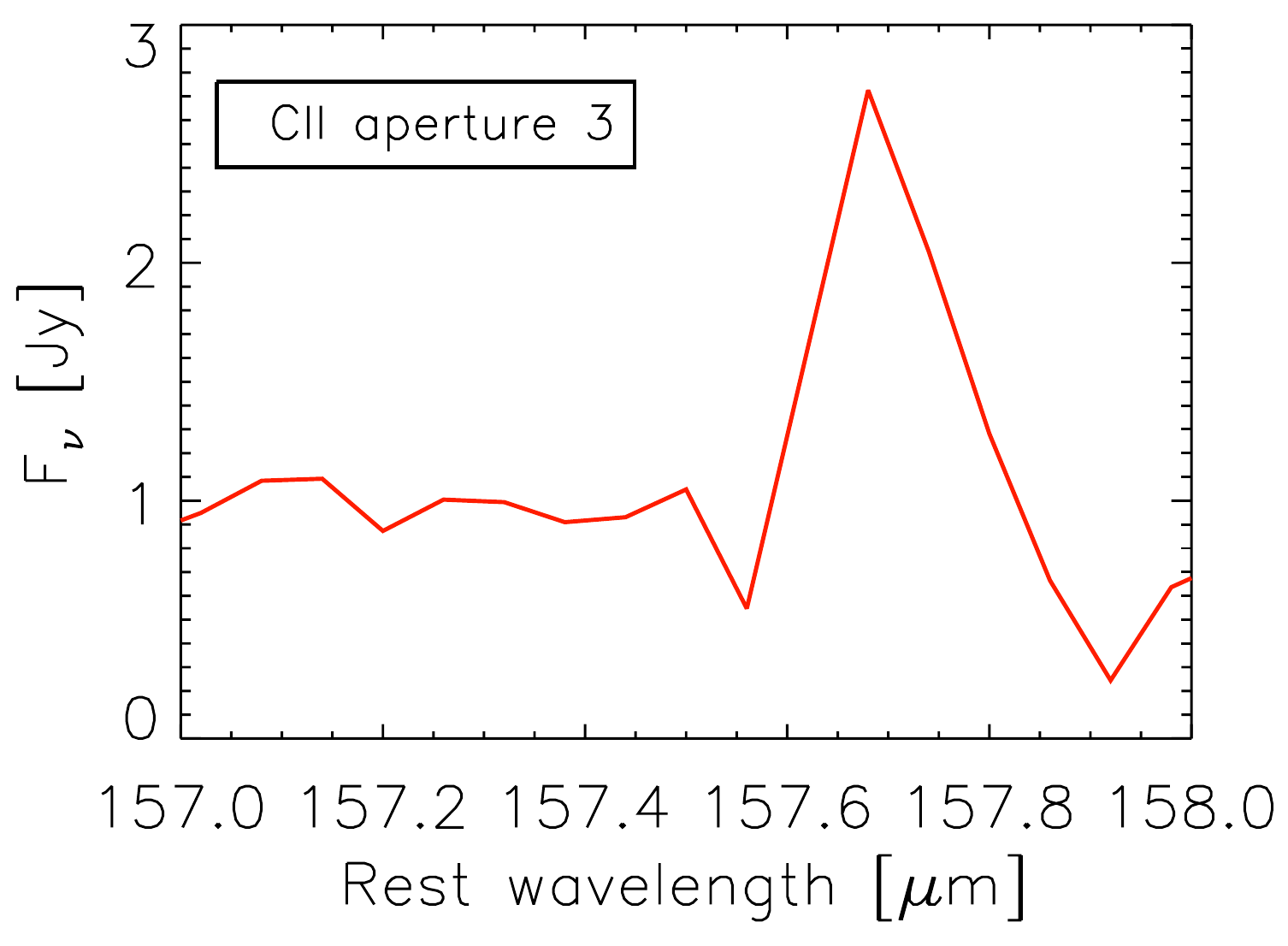} 
\includegraphics[width=35.5mm]{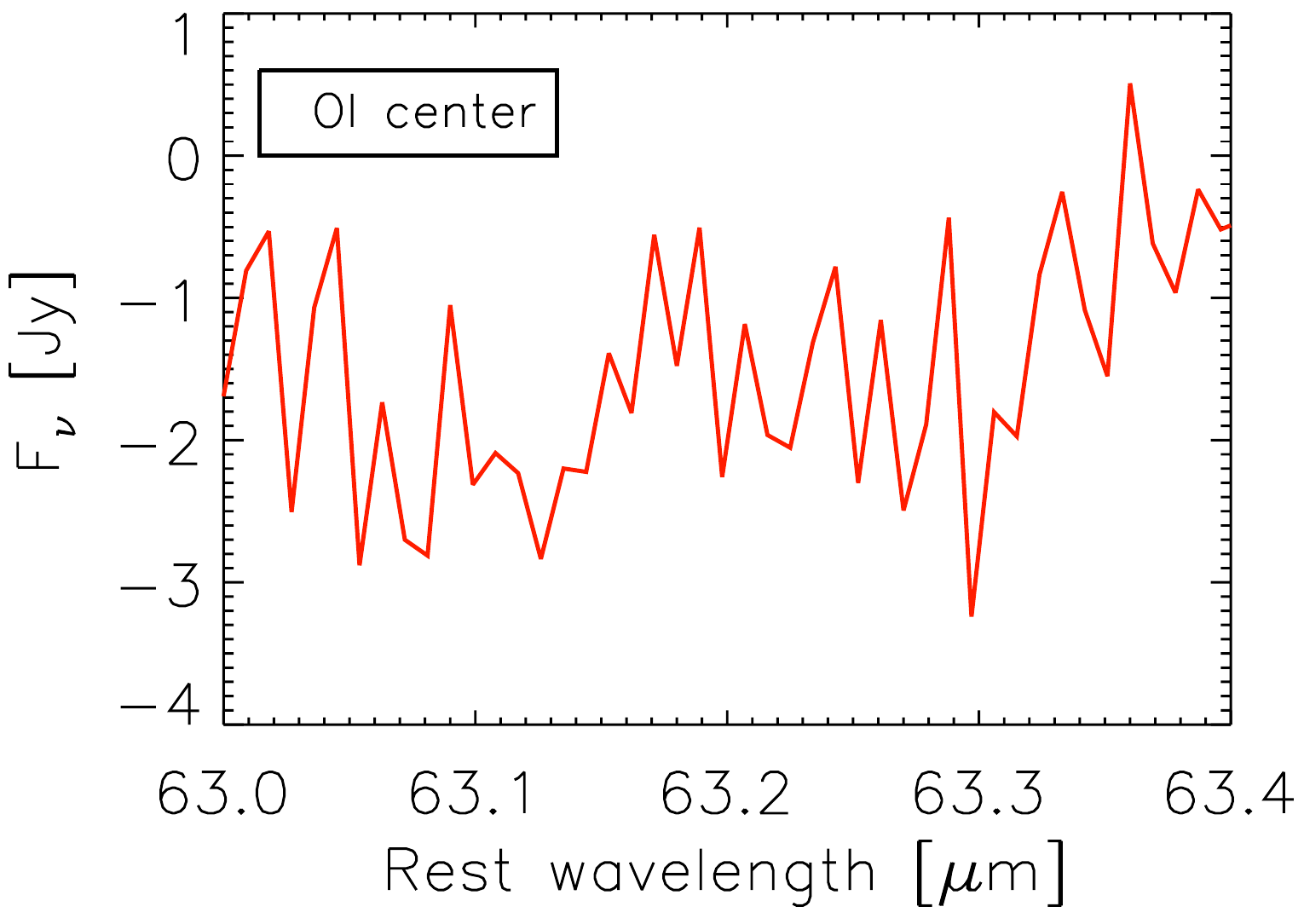} 
\includegraphics[width=35.5mm]{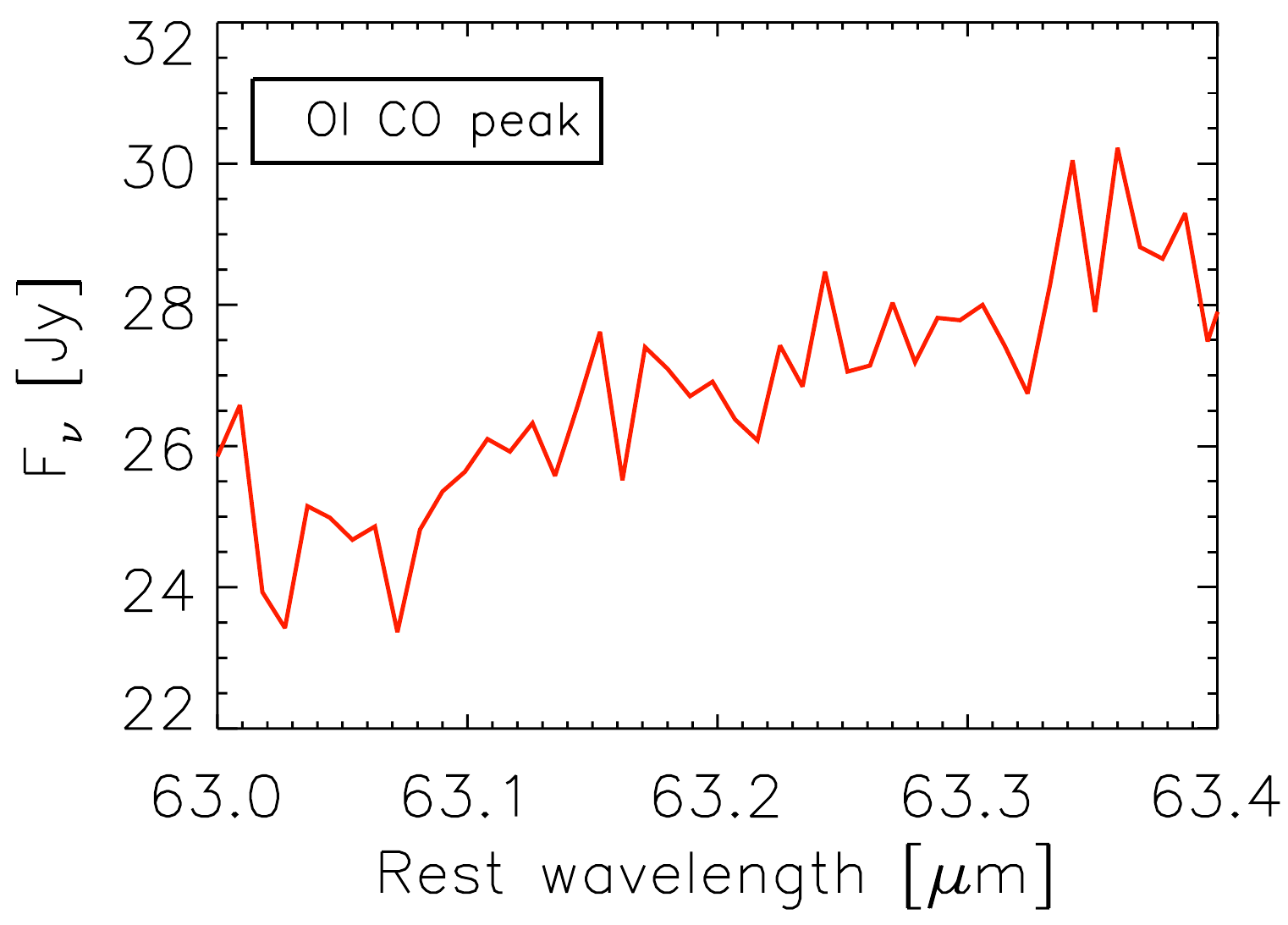} 
\caption{PACS spectral line plots. First three panels show the [C{\sc{ii}}] line emission summed over the individual apertures shown in Figure \ref{cii_map} and described in Table \ref{ciiflux}, respectively. The last two panels show the non-detection of the [O{\sc{i}}] line, with the emission added up within the observed regions covering the center and the CO peak in NGC\,205, respectively.}
\label{speclines}
\end{figure*}

From our PACS spectroscopy observations, we detected [C{\sc{ii}}] line emission in the center of NGC\,205 (see Figure \ref{cii_map}, left panel), while the [O{\sc{i}}] line was not detected in either of the two covered regions in NGC\,205 (see Figure \ref{speclines}). 
The faint [C{\sc{ii}}] emission in NGC\,205 originates mainly from the nuclear region, with the peak intensity residing from a dust cloud west of the nucleus. From the gas and dust contours in Figure \ref{cii_map} (right panel) it becomes evident that the brightest [C{\sc{ii}}] emission region is located at the boundary of the most massive H{\sc{i}} and molecular clouds in the center of NGC\,205. The spatial correlation with the CO reservoir in the center of NGC\,205 and the surrounding star formation regions (see contours of hot dust emission from MIPS 24 $\mu$m data, Figure \ref{cii_map}, right panel, blue, dashed-dotted curves) suggests that the [C{\sc{ii}}] emission in NGC\,205 originates from photodissociation regions in the outer layers of molecular cloud structures.
Although the [C{\sc{ii}}] emission in galaxies might have a significant contribution from ionized media, the unavailability of ionized gas tracers (e.g. [N{\sc{ii}}], [O{\sc{iii}}]) impede a direct quantification of the [C{\sc{ii}}] emission from ionized regions. The non-detection of H$\alpha$ line emission \citep{1997ApJ...476..127Y} seems to point towards a low fraction of ionized gas in the ISM of NGC\,205.

To determine the intensity of the [C{\sc{ii}}] line emission in NGC\,205, we performed aperture photometry within ellipses matching the shape of prominent [C{\sc{ii}}] emission regions in NGC\,205 (see Figure \ref{cii_map}, left panel). The spectral line emission within each emission region is shown in Figure \ref{speclines} (first three panels). The fluxes within those three apertures are summarized in Table \ref{ciiflux}.
From equation 1 in \citet{1993ApJ...407..579M}, we can derive the column density of atomic hydrogen, when adopting a C$^{+}$ abundance per hydrogen atom of $X_{\text{C}^{+}}$ $=$ 1.4 $\times$ 10$^{-4}$ \citep{1996ARA&A..34..279S} and $n_{\text{crit}}$ $\sim$ 2.7 $\times$ 10$^{3}$ cm$^{-3}$. When assuming $X_{\text{ISRF}}$ $\sim$ 3 (or thus $G_{0}$ $\sim$ 3) for the strength of the ISRF field (see Section \ref{Dust.sec}) and an average density $n_{\text{H}}$ $\sim$ 10$^{4}$ cm$^{-3}$, we estimate a surface density temperature of T $\sim$ 40 K from the PDR models in \citet{1999ApJ...527..795K}. Inserting those values in equation (1.) from \citet{1993ApJ...407..579M}, we obtain average column densities for the three apertures within the range 2.2 $\times$ 10$^{19}$ cm$^{-2}$ $\leq$ $N_{\text{H}}$ $\leq$ 3.7 $\times$ 10$^{19}$ cm$^{-2}$.
Summing over all three apertures, the total atomic gas mass for the [C{\sc{ii}}] emitting regions in NGC\,205 is derived to be $M_{\text{g}}$ $\sim$ 1.54 $\times$ 10$^{4}$ M$_{\odot}$. Since this value is negligible compared to the total gas mass $M_{g}$ $\sim$ 6.9 $\times$ 10$^{6}$ M$_{\odot}$ derived from CO(1-0) observations (see Section \ref{gasobs.sec}), we conclude that the molecular gas reservoir is probably well-traced by the CO lines in the low metallicity ISM of NGC\,205. 
This argument is furthermore supported by the relatively low $L_{\text{[CII]}}$-to-$L_{\text{CO(1-0)}}$ line intensity ratio $\sim$ 1850 within the central CO(1-0) pointing from \citet{1998ApJ...499..209W} (see green aperture overlaid on Figure \ref{cii_map}). This line intensity ratio is substantially lower than the values observed in star-forming dwarf galaxies ranging from 4000 to 80000 \citep{2000NewAR..44..249M,2010A&A...518L..57C}.

Based on PDR models for values $X_{\text{ISRF}}$ $\sim$ 3 and $n_{\text{H}}$ $\sim$ 10$^{4}$ cm$^{-3}$ in the center of NGC\,205, we would expect [C{\sc{ii}}] line intensities of a few 10$^{-6}$ erg cm$^{-2}$ s$^{-1}$ sr$^{-1}$ and $L_{\text{[OI]}}$-to-$L_{\text{[CII]}}$ line intensity ratios close to 0.5 (see Figure 3 and 4 from \citealt{1999ApJ...527..795K}). With a peak [C{\sc{ii}}] line intensity of 3 $\times$ 10$^{-6}$ erg cm$^{-2}$ s$^{-1}$ sr$^{-1}$ and an upper limit of $L_{\text{[OI]}}$-to-$L_{\text{[CII]}}$ $<$ 1.75, the observed [C{\sc{ii}}] line emission in the center of NGC\,205 is in agreement with PDR models.

The lack of any bright [C{\sc{ii}}] emission from other regions in NGC\,205 might be an indication for the strength of the radiation field being insufficient to photo-dissociate CO. This argument is supported by the pixel-by-pixel analysis in Section \ref{pixelbypixel.sec}, which indicates the stronger radiation field in the brightest [C{\sc{ii}}] region compared to the rest of the galaxy (see Figure \ref{pixelbypixel}, left panel). 
Alternatively, a large reservoirs of photodissociated CO molecules might be present in the outer layers of gas clouds, where a deficiency of ionizing photons might impede the formation of C$^{+}$ atoms. A lack of ionizing photons is also supported by the non-detection of H$\alpha$ in this galaxy \citep{1997ApJ...476..127Y}. According to this latter scenario, we would expect the majority of dissociated carbon to be locked in [C{\sc{i}}] rather than C$^{+}$. 
 
\section{Discussion}
\label{Discussion.sec}

\subsection{Missing ISM mass problem: revised}
\label{Problem.sec}
From our \textit{Herschel} observations of the dust continuum and far-infrared fine structure lines [C{\sc{ii}}] and [O{\sc{i}}] and JCMT CO(3-2) data, we are able to revisit the missing ISM mass problem in NGC\,205. 
We estimate an atomic gas mass of 1.54 $\times$ 10$^{4}$ M$_{\odot}$ associated with the [C{\sc{ii}}] emitting PDRs in NGC\,205.
From the CO(3-2) emitting regions in the northern part of NGC\,205, we could derive a molecular gas mass of  $M_{\text{H}_{2}}$ $\sim$ 1.3 $\times$ 10$^5$ M$_{\odot}$. In comparison with the $M_{\text{H}_{2}}$ $\sim$ 6.9 $\times$ 10$^5$ M$_{\odot}$ obtained from pointed CO(1-0) observations \citep{1998ApJ...499..209W} covering the main CO(1-0) emission regions in the north of the galaxy, our CO(3-2) measurements indicate a low fraction of dense molecular gas in the diffuse ISM of NGC\,205, where star formation is currently only occurring spontaneously in localized dense clouds. 
Including H{\sc{i}} observations (4.0 $\times$ 10$^5$ M$_{\odot}$) and scaling the sum of the molecular and atomic hydrogen mass by a factor $\sim$ 1.4 to include heavier elements, we obtain a total gas mass of $M_{g}$ $\sim$ 0.7-1.5 $\times$ 10$^6$ M$_{\odot}$ from CO(3-2) or CO(1-0) observations, respectively.

Alternatively, we probed the ISM content through the galaxy's dust continuum emission.
This approach avoids introducing uncertainty factors arising from the $X_{\text{CO}}$ conversion factor, but additional errors on the gas mass estimates result from the SED fitting procedure to calculate the dust mass and, even more importantly, the assumption on the value for the gas-to-dust ratio. Since it has been shown that the gas-to-dust ratio in metal-poor galaxies deviates from the Galactic ratio ($\sim$ 160, \citealt{2004ApJS..152..211Z}), a gas-to-dust fraction of $\sim$ 400 \citep{2008ApJ...672..214G} is considered more realistic for the central ISM in NGC\,205 in view of the recent star formation episode occurring in those inner regions. This value is obtained from an extrapolation of the dust evolution model in \citet{2008ApJ...672..214G}, when assuming little or no dust destruction and a first-order trend of the gas-to-dust ratio with metallicity (Z $\sim$ 0.3 Z$_{\odot}$). 
Relying on this gas-to-dust ratio of $\sim$ 400, the total dust mass detected from our \textit{Herschel} observations corresponds to a gas mass $M_{g}$ = $\sim$ 4-7 $\times$ 10$^6$ M$_{\odot}$.

Upon comparison with the theoretical gas content in the range [1.3 $\times$ 10$^7$ M$_{\odot}$, 4.8 $\times$ 10$^8$ M$_{\odot}$] (see Section \ref{Theory.sec}), 
gas mass estimates from either dust continuum (4-7 $\times$ 10$^6$ M$_{\odot}$) or H{\sc{i}}+CO(1-0)+[C{\sc{ii}}] (1.5 $\times$ 10$^6$ M$_{\odot}$) observations both confirm the missing ISM mass problem in NGC\,205. The lower ISM mass inferred from direct gas observations (1.5 $\times$ 10$^6$ M$_{\odot}$) in comparison with the indirect gas estimates from dust continuum observations (4-7 $\times$ 10$^6$ M$_{\odot}$) most likely results from the poor coverage of molecular gas tracers in NGC\,205 and/or the uncertainty on the assumed gas-to-dust fraction (i.e. a gas-to-dust ratio closer to the Galactic value would bring the ISM masses inferred from H{\sc{i}}+CO(1-0)+[C{\sc{ii}}] and dust continuum observations in better agreement).
With the JCMT CO(3-2) and previous CO(1-0) and CO(2-1) observations mainly covering the northern part of the galaxy, the lack of knowledge about the molecular gas content for a large southern region in NGC\,205, where the H{\sc{i}} emission dominates, prevents us from determining the entire molecular gas content. If we assume a similar molecular-to-atomic gas mass ratio for the southern part of NGC\,205 as observed in the northern part, the gas component in NGC\,205 could be twice as massive than measured by current CO observations.
The observed peak in H{\sc{i}} (4 $\times$ 10$^{20}$ cm$^{-2}$ or 3.2 M$_{\odot}$ pc$^{-2}$) and dust (0.1 M$_{\odot}$ pc$^{-2}$) column density would imply a H{\sc{i}}-to-dust ratio of $\sim$ 32 in the south of NGC\,205. However, a more realistic gas-to-dust ratio would require a large molecular gas reservoir residing in those southern areas. 
This argumentation is also confirmed by the H$_{2}$ column densities of 1.8 and 2.4 $\times$ 10$^{20}$ cm$^{-2}$ measured from two JCMT CO(2-1) pointings near the southern peak in H{\sc{i}}, when assuming a CO(2-1)-to-CO(1-0) line ratio of $\sim$ 0.9 $\pm$ 0.2 as found by \citet{1996ApJ...464L..59Y} in a spatially resolved molecular cloud.

In addition to this unexplored molecular gas reservoir in the southern part of the galaxy, a substantial fraction of the gas component might remain undetected by current observations, either because a fraction of the gas is locked in hot X-ray or ionized gas haloes (similar to the heated gas returned by evolved stars in giant massive ellipticals) and/or the present molecular gas is traced neither by CO nor [C{\sc{ii}}] molecules.
Alternatively, we could invoke an internal/environmental mechanism solely removing the gas from NGC\,205 and leaving the dust unharmed to explain the higher gas masses inferred from dust continuum observations. In view of the good correlation found for the gas and dust component, there is however no reason to assume a higher inertia for the dust and thus to imply that gas particles are more easily transported out of the gravitationally bound regions in NGC\,205.

In the next paragraphs, we discuss possible explanations for the inconsistency between the observed ISM content (either from H{\sc{i}}+CO(1-0)+[C{\sc{ii}}] or dust continuum observations) in NGC\,205 and theoretical predictions, such as non-standard conditions for the SFE or initial mass function (IMF) or environmental processes able to remove part of the ISM during the last episode of star formation in NGC\,205.

\subsection{Explaining the inconsistency regarding the ISM mass}
\label{Explanation.sec}

\subsubsection{Non-standard conditions}
Non-standard conditions such as a top-heavy IMF (more massive stars are produced, requiring less input gas mass) or a higher star formation efficiency ($>$ 10 $\%$ of the gas is converted into stars) could be invoked to explain the missing ISM in NGC\,205.

Harmonizing the observed ISM content with the theoretical predictions would require an IMF deviating significantly from the general assumptions and/or an increase in the SFE from 10$\%$ to at least 65$\%$.
Although such a top-heavy IMF or increased SFE was thought to be present in some ULIRGS or starbursts (e.g. M82, \citealt{1980ApJ...240...60K}), the non-standard conditions in those galaxies are owing to the presence of high density gas, introducing a different mode of star formation \citep{2010ApJ...714L.118D}. With a SFR $\sim$ 7.0 $\times$ 10$^{-4}$ M$_{\odot}$ yr$^{-1}$ \citep{2009A&A...502L...9M} during the last starburst episode, NGC\,205 does not mimic the typical star formation activity ($>$ 10 M$_{\odot}$ yr$^{-1}$) in those starburst galaxies. Also under less extreme circumstances, local variations in the SFE \citep{2011A&A...533A..19B} and perturbations at the upper mass end of the IMF \citep{Meurer} have been claimed. However, the latter IMF variations have been questioned on its turn invoking poor extinction corrections and a variable SFR in those galaxies \citep{2009ApJ...706.1527B, Weisz}. 
Relying on recent results from \citet{2011arXiv1111.1925M} and \citet{Boylan}, finding no evidence for extreme star formation efficiencies in star-forming dwarf galaxies and therefore rather supporting a moderate SFE of 10-20$\%$ or less, we argue that non-standard conditions (top-heavy IMF and/or increased SFE) are not likely to occur in NGC\,205.  and we should invoke efficient mechanisms of gas removal to explain the low observed gas content in NGC\,205.

\subsubsection{Supernova feedback}

Feedback from supernovae \citep{1986ApJ...303...39D, 1992MNRAS.255..346B, 1994ApJ...431..598D} or potentially an AGN are capable of removing a significant amount of gas from NGC\,205. 
Relying on the absence of any LINER or Seyfert diagnostics \citep{1995ApJS...98..477H}, we argue that the effect of AGN feedback in NGC\,205 is currently negligible. 
Based on the 3$\sigma$ upper limit of 3.8 $\times$ 10$^4$ M$_{\odot}$ for the mass of the SMBH in NGC\,205 from studies of the stellar kinematics based on deep HST images \citep{2005ApJ...628..137V}, we believe the role of an AGN was also limited throughout the recent history of NGC\,205.
On the contrary, supernova feedback and/or stellar winds could be responsible for expelling gas from the central regions, during the last episode of star formation. Examining the effectiveness of SN II feedback during the last episode of star formation,
\citet{1998ApJ...499..209W} confirm that supernova winds or blasts were capable of removing a significant amount of gas from the central regions of NGC\,205. Indeed, applying the dwarf galaxy model for $M_{galaxy}$ $\sim$ 10$^9$ M$_{\odot}$ from \citet{1994ApJ...431..598D}, a minimum energy of $\sim$ 10$^{55}$  f$_{gas}$ ergs would be required to expel most of the gas from the inner regions. Following the burst mass $\sim$ 7 $\times$ 10$^5$ M$_{\odot}$ reported in \citet{1990ApJ...364...87W}, the gas fraction is found to be f$_{gas}$ $\sim$ 0.007, when assuming a 10$\%$ star formation efficiency. This would require an energy of at least 7 $\times$ 10$^{52}$ ergs to remove gas. This level is easily achieved with the energy $\sim$ 6 $\times$ 10$^{54}$ ergs released by the SNe II associated with the last burst \citep{1998ApJ...499..209W}. Keeping in mind the clumpy gas distribution and the offset of the local H{\sc{i}} column density peaks from the locations of young stars \citep{1997ApJ...476..127Y} in addition to those theoretical arguments, we argue that both supernova feedback and/or stellar winds have likely disturbed the ISM of NGC\,205. Energy feedback from supernovae has furthermore been proven to have an important share in the formation of dwarf spheroidal systems \citep{2009ApJS..182..216K}.

Despite theoretical arguments supporting a history of violent supernova explosions, a previous attempt to detect supernova remnants in NGC\,205 failed ($F_{\nu}$ $<$ 0.06 mJy, \citealt{2007AJ....134.2148L}).
Relying on the low radio continuum detection rate for low surface brightness dwarfs \citep{1994AJ....108..446H, 2007AJ....134.2148L}, it might not be surprising that the detection of supernova remnants at 20 cm was unsuccessful. The majority of the 20 cm emission is thought to arise from synchrotron emission originating from electrons accelerated in the expanding shells of Type II and Type Ib supernova remnants.
Due to the increased cosmic ray diffusion timescales \citep{1992A&A...255...49K, 2006ApJ...651L.111M} (i.e. electrons are escaping more easily from the ISM) for galaxies with a low star formation activity (SFR $\leq$ 0.2 M$_{\odot}$ yr$^{-1}$), those objects become radio quiet in short timescales.
Although violent supernova explosions might have occurred in the past, the associated radio emission is likely smoothed out in the low potential well of NGC\,205. Also in other wavelength domains the detection of any supernova remnants will be challenging. While the star formation is known to be active up to at least 60 Myr ago, a typical age $\sim$ 10 Myr for a SNII progenitor and a SNR lifetime $\sim$ 25000 years implies that it is difficult to detect any remaining evidence of supernova remnants in NGC\,205.
 
\subsubsection{Environmental interactions}
Galaxy harassment \citep{1996Natur.379..613M}, starvation \citep{1980ApJ...237..692L} and viscous stripping \citep{1982MNRAS.198.1007N} are important transformation processes for galaxies in clusters, whereas tidal stirring \citep{2001ApJ...559..754M} and galaxy threshing \citep{2001ApJ...552L.105B} are considered responsible for the formation of dwarf spheroidals and ultra-compact dwarfs in the low density environment of groups.
For the formation of NGC\,205, a combination of ram pressure and tidal stripping would be more likely since those processes are capable of transforming gas-rich dwarf galaxies into dwarfs with a blue central core when passing through the halo of a galaxy of the same size as the Milky Way \citep{2006MNRAS.369.1021M}. Tidal or gravitational interactions are also considered to be the main formation mechanism for spheroidal and lens-shaped galaxies in groups \citep{2009ApJS..182..216K,2011MNRAS.415.1783B,matt}. In a similar fashion, ram pressure stripping is found to form dwarf elliptical galaxies with blue nuclei in the Virgo cluster \citep{2008ApJ...674..742B}.

Several observations are indicative for the tidal influence of M31 on its companion NGC\,205: a twist in the elliptical isophotes \citep{1973ApJ...182..671H, 2002AJ....124..310C}, a stellar arc of blue metal-poor red giant branch stars northwest of M31 \citep{2001Natur.412...49I,2004MNRAS.351L..94M}, a tidal debris of C stars to the west of NGC\,205 \citep{2003AJ....125.3037D}, a stripped H{\sc{i}} cloud 25$\arcmin$ southwest of NGC\,205 which overlaps in velocity with the galaxy \citep{2004ApJ...601L..39T}, peculiar H{\sc{i}} morphology and kinematics \citep{1997ApJ...476..127Y}, stars moving in the opposite direction with respect to the rotation of the main stellar body \citep{2006AJ....131..332G, 2006MNRAS.369.1321D} and an optical tidal tail extending at least 17$\arcmin$ southwards from the galaxy's center \citep{2010IAUS..262..426S}. 
This latter stellar tail coincides with the tentative dust tail detected in our SPIRE data (see Section \ref{Distribution.sec}) and, thus, might be an indication for the removal of dust from NGC\,205 as a result from the tidal influence of M31.
Besides the tidal features characterizing NGC\,205, M31 is also affected by the tidal influence exerted by NGC\,205. While the origin of two off-centre spiral rings is probably attributable to a head-on collision with M32 \citep{2006Natur.443..832B}, the warped structure in the outer H{\sc{i}} disk of M31 \citep{1986PASJ...38...63S} and a distortion of the spiral structure in the disk \citep{2006ApJ...638L..87G} are likely caused to some extent by a two-body interaction between M31 and NGC\,205. 

Although those observations confirm the tidal influence of M31 on the outer regions of NGC\,205, it does not provide an explanation for the removal of gas from the inner regions, where the last episode of star formation took place and we would expect to find the left-over gas reservoir.
Hydrodynamical simulations suggest that the gaseous component might be disrupted and partly removed even within the tidal radius (4$\arcmin$, \citealt{2006AJ....131..332G}) without any indications for stellar bridges and tails at these radii \citep{1985A&A...144..115I}, but whether it is the case for NGC\,205 is difficult to inquire due to the uncertainties regarding its orbit around M31 \citep{2008ApJ...683..722H}.
However, in combination with supernova feedback expelling the gas from the inner regions, tidal interactions might strip the expelled gas from the outer regions.

Whether or not responsible for the removal of gas from the inner regions, a tidal encounter with a companion closer than 100-200 kpc is able to trigger SF through shocks in the disk of a typical dwarf galaxy when moving on a coplanar prograde orbit \citep{2004MNRAS.349..357B}. With a line-of-sight distance between the Andromeda galaxy and NGC\,205 of $\sim$ 39 kpc \citep{2005MNRAS.356..979M} and the likely prograde trajectory of NGC\,205 toward its parent galaxy, M31 \citep{2006AJ....131..332G, 2008ApJ...683..722H}, tidal interactions might have given the initial start for the last episode of star formation. 
Furthermore, the most recent episodes of star formation during the last gigayear seem correlated with the orbit about M31 \citep{2003ApJ...597..289D}, which provide additional evidence for the tidal triggering of star formation in NGC\,205.

\subsection{Comparison to other galaxies}
\label{Compare.sec}
Two other dwarf companions of M31, NGC\,185 and NGC\,147, are thought to have a star formation history comparable to NGC\,205 based on their nearly identical optical appearances (Holmberg diameters, $B$ - $V$ colours, average surface brightnesses), mass-to-light ratios ($M$/$L$)$_{B}$ $\sim$ 4 M$_{\odot}$/L$_{\odot,B}$  \citep{2006MNRAS.369.1321D} and specific frequencies of C stars \citep{2005AJ....130.2087D}, implying similar fractions of gas and dust which have been turned into stars in the past.
Interestingly, both objects also feature a missing ISM mass problem \citep{1998ApJ...507..726S}.

In correspondence to the missing ISM mass problem in NGC\,205, tidal interactions and supernova feedback are also believed to have influenced the ISM content in NGC\,185 and NGC\,147.
The tidal influence of M31 ($D$ = 785 $\pm$ 25 kpc, \citealt{2005MNRAS.356..979M}) is considered negligible at distances of $D$ = 616 $\pm$ 26 and 675 $\pm$ 27 kpc \citep{2005MNRAS.356..979M} for NGC\,185 and NGC\,147, respectively, compared to NGC\,205 ($D$ = 824 $\pm$ 27 kpc, \citealt{2005MNRAS.356..979M}). However, both galaxies are thought to form a gravitationally bound pair \citep{1998AJ....116.1688V} and might have tidally interacted in the past.
Aside from the possible occurrence of tidal interactions, indications for supernova remnants are found in NGC\,185 \citep{1984ApJ...281L..63G, 1997ApJ...476..127Y, 2007AJ....134.2148L} and the current ISM content in this galaxy ( $\sim$ 7.3 $\times$ 10$^5$ M$_{\odot}$) resembles the estimated mass returned to the ISM by planetary nebulae ($\sim$ 8.4 $\times$ 10$^5$ M$_{\odot}$, \citealt{1998ApJ...507..726S}).
However, the lack of ISM in NGC\,147 remains a puzzling feature in this evolutionary framework.

\section{Conclusions}
\label{Conclusions.sec}

This work reports on \textit{Herschel} dust continuum, [C{\sc{ii}}] and [O{\sc{i}}] spectral line and JCMT CO(3-2) observations for NGC\,205, the brightest early-type dwarf satellite of the Andromeda galaxy.
While direct gas observations (H{\sc{i}}+CO(1-0): 1.5 $\times$ 10$^6$ M$_{\odot}$) have proven to be inconsistent with theoretical predictions of the current gas content in NGC\,205 ($>$ 10$^7$ M$_{\odot}$), we could revise the missing ISM mass problem based on new gas mass estimates (CO(3-2), [C{\sc{ii}}], [O{\sc{i}}]) and an indirect measurement of the ISM content probed through \textit{Herschel} dust continuum observations taken in the frame of the VNGS and HELGA projects. 

SED fitting to the FIR/submm fluxes results in a total dust mass $M_{\text{d}}$ $\sim$ 1.1-1.8 $\times$ 10$^4$ M$_{\odot}$ at an average temperature $T_{d}$ $\sim$ 18-22 K. Based on Herschel data, we can also exclude the presence of a massive cold  dust component ($M_{\text{d}}$ $\sim$ 5 $\times$ 10$^5$ M$_{\odot}$, $T_{d}$ $\sim$ 12 K), which was suggested based on millimeter observations from the inner 18.4$\arcsec$. 
When assuming a metal abundance $Z$ $\sim$ 0.3 Z$_{\odot}$ and a corresponding gas-to-dust ratio $\sim$ 400, a gas mass $M_{g}$ $\sim$ 4-7 $\times$ 10$^6$ M$_{\odot}$ is probed indirectly through Herschel dust continuum observations.

The non-detection of [O{\sc{i}}] and the relatively low $L_{\text{[CII]}}$-to-$L_{\text{CO(1-0)}}$ line intensity ratio ($\sim$ 1850) imply that the molecular gas phase is well traced by CO molecules. From CO(3-2) observations of the northern part of the galaxy, we infer a new molecular gas mass estimate $M_{\text{H}_{2}}$ $\sim$ 1.3 $\times$ 10$^{5}$ M$_{\odot}$, implying that the molecular clouds in NGC\,205 is mostly very diffuse. 

New gas mass estimates from dust continuum and CO(3-2) line observations both confirm the missing ISM mass problem, i.e. an inconsistency between theoretical predictions for the ISM mass and observations. 
In an attempt to explain the deficiency in the ISM in the inner regions of NGC\,205, we claim that efficient supernova feedback capable of expelling gas/dust from the inner, star-forming regions of NGC\,205 to the outer regions and/or tidal interactions with M31 stripping the gas/dust component from the galaxy provide the best explanation for the removal of a significant amount of the ISM from NGC\,205.  
However, if supernova feedback is found to lack responsibility for the removal of gas/dust from the inner regions, we might have to reconsider the importance of tidal interactions on the gaseous component in a galaxy within the tidal radius and/or revise the parameters characterizing the orbit of NGC 205 in its approach toward M31.

\section*{Acknowledgements}
PACS has been developed by a consor-
tium of institutes led by MPE (Germany) and in-
cluding UVIE (Austria); KU Leuven, CSL, IMEC
(Belgium); CEA, LAM (France); MPIA (Germany);
INAF- IFSI/OAA/OAP/OAT, LENS, SISSA (Italy);
IAC (Spain). This development has been supported by
the funding agencies BMVIT (Austria), ESA-PRODEX
(Belgium), CEA/CNES (France), DLR (Germany),
ASI/INAF (Italy), and CICYT/MCYT (Spain). SPIRE
has been developed by a consortium of institutes led by
Cardiff University (UK) and including Univ. Lethbridge
(Canada); NAOC (China); CEA, OAMP (France); IFSI,
Univ. Padua (Italy); IAC (Spain); Stockholm Ob-
servatory (Sweden); ISTFC and UKSA (UK); and Cal-
tech/JPL, IPAC, Univ. Colorado (USA). This devel-
opment has been supported by national funding agen-
cies: CSA (Canada); NAOC (China); CEA, CNES,
CNRS (France); ASI (Italy); MCINN (Spain); Stock-
holm Observatory (Sweden); STFC (UK); and NASA
(USA). HIPE is a joint development by the Herschel
Science Ground Segment Consortium, consist-
ing of ESA, the NASA Herschel Science Center and
the HIFI, PACS and SPIRE consortia.

This work was made possible by the facilities of the Shared Hierarchical 
Academic Research Computing Network (SHARCNET:www.sharcnet.ca) and 
Compute/Calcul Canada.

The research of C. D. W. is supported by grants from rom the Canadian Space Agency and the Natural Sciences and Engineering Research Council of Canada.

MB, JF, IDL and JV acknowledge the support of the Flemish Fund for Scientific Research (FWO-Vlaanderen),
in the frame of the research projects no. G.0130.08N and no. G.0787.10N .

GG is a postdoctoral researcher of the FWO-Vlaanderen (Belgium).

\bsp

\label{lastpage}

\end{document}